\documentclass[pre,twocolumn,preprintnumbers,amsmath,amssymb,nofootinbib,floatfix]{revtex4}

\usepackage{graphicx,bm}

\makeatletter
\def\graphicscale{\twocolumn@sw{0.3}{0.4}}
\def\graphicthreescale{\twocolumn@sw{0.3}{0.4}}

\begin{document}

\title{Three-dimensional monopole-free CP$^{N-1}$ models }

\author{Andrea Pelissetto}
\affiliation{Dipartimento di Fisica dell'Universit\`a di Roma Sapienza
        and INFN Sezione di Roma I, I-00185 Roma, Italy}

\author{Ettore Vicari} 
\affiliation{Dipartimento di Fisica dell'Universit\`a di Pisa
        and INFN Largo Pontecorvo 3, I-56127 Pisa, Italy}

\date{\today}

\begin{abstract}
We investigate the phase diagram and the nature of the phase
transitions of three-dimensional monopole-free CP$^{N-1}$ models,
characterized by a global U($N$) symmetry, a U(1) gauge symmetry,
and the absence of monopoles.  We present numerical analyses based on
Monte Carlo simulations for $N=2$, 4, 10, 15, and 25.  We observe a
finite-temperature transition in all cases, related to the
condensation of a local gauge-invariant order parameter.  For $N=2$ we
are unable to draw any definite conclusion on the nature of the
transition.  The results may be interpreted in terms of  either a weak
first-order transition or of a continuous transition with anomalously
large scaling corrections.  However, the results allow us to exclude
that the transition belongs to the O(3) vector
universality class, as it occurs in the standard three-dimensional
CP$^{1}$ model without monopole suppression. For $N=4,10,15$, the
transition is of first order, and significantly weaker than that
observed in the presence of monopoles. For $N=25$ the results are
consistent with a conventional continuous transition. We compare our
results with the existing literature and with the predictions of
different field-theory approaches.  They are consistent with
the scenario in which the model undergoes continuous transitions for
large values of $N$, including $N=\infty$, in agreement with analytic
large-$N$ calculations for the $N$-component Abelian-Higgs model.
\end{abstract}

\maketitle


\section{Introduction}
\label{intro}

Models of scalar fields with U(1) gauge symmetry and U($N$) global
symmetry have been extensively studied with the purpose of identifying
the nature of their different phases and transitions. They emerge as
effective theories of superconductors and superfluids and of quantum
SU($N$) antinferromagnets
~\cite{RS-90,TIM-05,TIM-06,Kaul-12,KS-12,BMK-13,NCSOS-15,WNMXS-17}.
In particular, three-dimensional (3D) classical models with $N=2$ are
supposed to describe the transition between the N\'eel and the
valence-bond-solid state in two-dimensional antiferromagnetic SU(2)
quantum systems
\cite{Sandvik-07,MK-08,JNCW-08,Sandvik-10,HSOMLWTK-13,HDKPS-13,PDA-13},
that represent the paradigmatic models for the so-called deconfined
quantum criticality \cite{SBSVF-04}.

In the last twenty years there has been an extensive discussion on the
nature of the transition occurring in this class of quantum models and
in their classical counterparts. It has been realized that
the nature of the transition  depends crucially on topological aspects, for
instance the Berry phase in the quantum case, the compact/noncompact
nature of the gauge fields and the presence/absence of monopoles in
the classical setting.  In this paper we wish to understand the role
that topological defects play in the simplest classical model with
U(1) gauge symmetry, the lattice CP$^{N-1}$ model. The fundamental
fields are complex $N$-component unit vectors ${\bm z}_{\bm x}$,
associated with the sites of a regular lattice --- we will consider
cubic lattices --- and U(1) gauge variables $\lambda_{{\bm
    x},\mu}=e^{i\theta_{{\bm x},\mu}}$ associated with the lattice
links.  The corresponding Hamiltonian is~\cite{RS-81,DHMNP-81,BL-81}
\begin{equation}
H = - N
\sum_{{\bm x}, \mu}
\left( \bar{\bm{z}}_{\bm x} 
\cdot \lambda_{{\bm x},\mu}\, {\bm z}_{{\bm x}+\hat\mu} 
+ {\rm c.c.}\right),
\label{hcpnla}
\end{equation}
where the sum is over all lattice sites ${\bm x}$ and directions $\mu$
($\hat{\mu}$ are the corresponding unit vectors).  The partition
function is
\begin{equation}
Z = \sum_{\{{\bm{z}}_{\bm x},\lambda_{{\bm x},\mu}\} } 
      e^{-\beta H} .
\end{equation}
The factor $N$ in the Hamiltonian (\ref{hcpnla}) is introduced for
convenience; with this definition, the large-$N$ limit is defined by
taking $N\to\infty$ keeping $\beta$ fixed.  One can easily check that
Hamiltonian (\ref{hcpnla}) is invariant under the global U($N$)
transformations
\begin{equation}
{\bm z}_{\bm x} \to U {\bm z}_{\bm x} ,\qquad U\in {\rm U}(N),
\label{unsym}
\end{equation}
and  the local U(1) gauge symmetry 
\begin{equation}
{\bm z}_{\bm x} \to e^{i\alpha_{\bm x}} {\bm z}_{\bm x} , \qquad
\lambda_{{\bm x},\mu} \to 
e^{i\alpha_{\bm x}} \lambda_{{\bm x},\mu}  e^{-i\alpha_{\bm x+\hat{\mu}}} .
\label{u1gausym}
\end{equation}
The model has a continuous transition for $N=2$ in the O(3)
universality class, while the transition is of first order for any
$N\ge 3$ \cite{PV-19-CP,PV-20-largeN}.  Note that the transition is
not continuous even for $N=\infty$, in disagreement with analytic
calculation \cite{DHMNP-81,PV-19-CP} performed for this model (see
Ref.~\cite{PV-20-largeN} for a discussion).

As we already mentioned, we expect the critical behavior to depend on
topological properties.  Topological defects like monopoles (or
hedgehogs) are supposed to be relevant in determining the phase
behavior.  For instance, the disordered phase and the corresponding
phase transition is absent in an O(3) vector model in which all
hedgehogs are suppressed \cite{LD-89,KM-93}, while a partial suppression
leads to a phase transition that appears different from the Heisenberg
one \cite{KM-93,MV-04}.  Analogously, the failure of the usual analytic
calculations in the large-$N$ limit for model (\ref{hcpnla}) has been
ascribed to the presence of topologically nontrivial configurations
that forbid the ordering of the gauge fields in the high-temperature
phase \cite{MS-90,PV-20-largeN}.

To explore the role that topological defects play in classical scalar
U(1) gauge systems, we consider the monopole-free CP$^{N-1}$
(MFCP$^{N-1}$) model. In this model the statistical average is
performed by summing only over the gauge-field configurations in which
monopoles are absent, where monopoles are defined using the De
Grand-Toussaint prescription \cite{DGT-80}.  The model we consider here 
is strictly related with the Abelian Higgs model with noncompact
gauge fields, which is often referred to as noncompact CP$^{N-1}$ model in the
literature on deconfined quantum criticality, see, e.g., 
Refs.~\cite{MV-04,NCSOS-15}, and, for $N=2$, to 
the O(3) model with hedgehog suppression discussed
in Refs.~\cite{LD-89,KM-93,MV-04}. They all share the same global symmetry
group and are characterized by the suppression of topological defects.

We consider 
different values of $N$, i.e. $N=2,4,10,15$ and 25.  In all cases, we
observe a finite-temperature transition associated with the local
order parameter
\begin{equation}
Q_{{\bm x}}^{ab} = \bar{z}_{\bm x}^a z_{\bm x}^b - {1\over N}
\delta^{ab},
\label{qdef}
\end{equation}
which is a gauge-invariant hermitian and traceless $N\times N$ matrix
that transforms as
\begin{equation}
Q_{{\bm x}} \to {U}^\dagger Q_{{\bm x}} \,{U}
\label{symmetry-U(N)}
\end{equation}
under the global U($N$) transformations (\ref{unsym}).  

We analyze the nature of the transition using finite-size scaling
(FSS) methods. In all cases, we observe that the suppression of
monopoles changes significantly the behavior of the system. For $N=2$
our results are definitely not consistent with an O(3) continuous
transition. Monopoles are essential to guarantee the Heisenberg
nature of the transition for $N=2$. We are however unable to establish
the order of the transition for the MFCP$^1$ model.  Our data are
consistent with a very weak first-order transition or with a
continuous transition with large scaling corrections.  For $N=4,10,15$
the transition is of first order, as in the model with monopoles, but
is significantly weaker.  Finally, for $N=25$, we observe a
continuous transition.  The latter result implies the
existence of a value $N_c$ such that the first-order transition
observed for $4 \le N \le 15$ turns into a continuous one as $N$
increases beyond $N_c$.  This leads us to conjecture that the
MFCP$^{N-1}$ model has a continuous transition in the large-$N$ limit,
as predicted by a perturbative analysis of the 
Abelian-Higgs field theory~\cite{DHMNP-81,MZ-03}.

The paper is organized as follows.  In Sec.~\ref{sec2} we define the
3D MFCP$^{N-1}$ model we consider, while in Sec.~\ref{sec3} we
define the basic observables that are determined in the Monte Carlo
simulations. The numerical results are presented in
Sec.~\ref{sec4}. In Sec.~\ref{sec4.1} we present the results for
$N=4,10,15$, which are all consistent with a first-order transition.
In Sec.~\ref{sec4.2} we discuss the results for $N=2$.  In spite of
the large simulated systems, we are unable to draw any conclusion on
the order of the transition. Finally, in Sec.~\ref{sec4.3} we discuss
the results for $N=25$, which are definitely consistent with a
continuous transition.  Finally, in Sec.~\ref{sec5} we summarize our
main results and compare them with the existing relevant literature.

\section{The lattice MFCP$^{N-1}$ model}
\label{sec2}

In our study we consider the CP$^{N-1}$ model with Hamiltonian
(\ref{hcpnla}) on a cubic lattice with periodic boundary conditions.
We define monopoles and antimonopoles using the De Grand-Toussaint
prescription \cite{DGT-80}.  In this approach one starts from the
noncompact lattice curl $\Theta_{{\bm x},\mu\nu}$ associated with each
plaquette
\begin{equation}
  \Theta_{{\bm x},\mu\nu} = 
    \theta_{{\bm x},\mu} + \theta_{{\bm x} + \hat{\mu},\nu} - 
    \theta_{{\bm x},\nu} - \theta_{{\bm x} + \hat{\nu},\mu},
\end{equation}
where $\theta_{{\bm x},\mu}$ is the phase associated with
$\lambda_{{\bm x},\mu}$, $\lambda_{{\bm x},\mu} = e^{i\theta_{{\bm
      x},\mu}}$.  Here $\mu$ and $\nu$ are the directions that
identify the plane in which the plaquette lies. Note that
$\Theta_{{\bm x},\mu\nu}$ is antisymmetric in $\mu$ and $\nu$, so that
we associate two different quantities that differ by
a sign with each plaquette. 
Let us now consider a closed lattice surface $S$ made of
elementary plaquettes. We associate $\Theta_{{\bm x},\mu\nu}$
with  each plaquette $P = ({\bm x},\mu\nu)$ , ordering $\mu$ and $\nu$ so that
the unit vector $\hat{\mu}\times \hat{\nu}$ points outward with
respect to the surface. It is then easy to verify that
\begin{equation}
   \sum_{P\in S} \Theta_{{\bm x},\mu\nu} = 0.
\label{Sumtheta}
\end{equation}
Indeed, with the chosen orientation of the plaquettes, each variable
$\theta_{{\bm x},\mu}$ [$({\bm x},\mu$) is a link belonging to $S$]
appears twice in the sum (\ref{Sumtheta}), with opposite sign; it
follows that all terms cancel, obtaining Eq.~(\ref{Sumtheta}). To
define monopoles, let us introduce the function
\begin{equation}
   m(x) = x - \left\lfloor x + 1/2 \right\rfloor.
\end{equation}
It satisfies $-1/2\le m(x) < 1/2$ and the relation $m(x) = x$ for any
$x$ in the interval $[-1/2,1/2[$. Moreover $m(x) - x$ is always an
    integer. We can now define the number of monopoles/antimonopoles
    within the surface $S$ as
\begin{equation}
N_{\rm mono}(S) = \sum_{P\in S} 
    m\left( {\Theta_{{\bm x},\mu}  \over 2\pi} \right).
\end{equation}
Because of the relation (\ref{Sumtheta}), $N_{\rm mono}(S)$ is always
an integer. Note that a nonvanishing number is only obtained if
$|\Theta_{{\bm x},\mu}| > \pi$ on some plaquettes. Thus, a finite density 
of monopoles is only observed 
in the disordered high-temperature phase, up to the critical point. 
In the low-temperature phase, only isolated pairs of a monopole and an
antimonopole are present. Their number decreases rapidly with
increasing $\beta$, since $\theta_{{\bm x},\mu} = 0$ (mod $2\pi$) on
all plaquettes for $\beta \to \infty$.

To define a monopole-free version of the CP$^{N-1}$ model, 
which we name MFCP$^{N-1}$, we restrict
our configuration space, considering only configurations for which
$N_{\rm mono}(C)=0$ on any elementary lattice cube.

\section{The observables} \label{sec3}

In our numerical study we consider cubic lattices of linear size $L$
with periodic boundary conditions.  We simulate the system using the
same overrelaxation algorithm we employed in our previous work
\cite{PV-19-AH3d,PV-20-largeN}.  It consists in a stochastic mixing of
microcanonical and standard Metropolis updates of the lattice
variables~\cite{footnoteMC}. The only difference is the addition of a
check: if the proposed move generates a monopole, the move is
rejected.

We compute the energy density and the specific heat, defined as
\begin{eqnarray}
E = {1\over N V} \langle H \rangle,\qquad
C ={1\over N^2 V}
\left( \langle H^2 \rangle 
- \langle H \rangle^2\right),
\label{ecvdef}
\end{eqnarray}
where $V=L^3$.  We consider correlations of the gauge invariant
operator $Q_{\bm x}^{ab}$ defined in Eq.~(\ref{qdef}).  Its two-point
correlation function is defined as
\begin{equation}
G({\bm x}-{\bm y}) = \langle {\rm Tr}\, Q_{\bm x}^\dagger  
Q_{\bm y} \rangle,  
\label{gxyp}
\end{equation}
where the translation invariance of the system has been taken into
account.  The susceptibility and the correlation length are defined as
\begin{eqnarray}
&&\chi =  \sum_{{\bm x}} G({\bm x}) = 
\widetilde{G}({\bm 0}), 
\label{chisusc}\\
&&\xi^2 \equiv  {1\over 4 \sin^2 (\pi/L)}
{\widetilde{G}({\bm 0}) - \widetilde{G}({\bm p}_m)\over 
\widetilde{G}({\bm p}_m)},
\label{xidefpb}
\end{eqnarray}
where $\widetilde{G}({\bm p})=\sum_{{\bm x}} e^{i{\bm p}\cdot {\bm x}}
G({\bm x})$ is the Fourier transform of $G({\bm x})$, and ${\bm p}_m =
(2\pi/L,0,0)$. In our FSS analysis we use renormalization-group invariant 
quantities. We consider
\begin{equation}
R_\xi = \xi/L
\end{equation}
and the Binder parameter
\begin{equation}
U = {\langle \mu_2^2\rangle \over \langle \mu_2 \rangle^2} , \qquad
\mu_2 = {1\over V^2} \sum_{{\bm x},{\bm y}} {\rm Tr}\, Q_{{\bm
    x}}^\dagger Q_{\bm y} .
\label{binderdef}
\end{equation}
We also consider the gauge-invariant vector
correlation function \cite{PV-19-AH3d,PV-20-largeN}
\begin{eqnarray} 
\label{Gv}
G_V(t,L) = {1\over 3 V} \sum_{{\bm x},\mu} \hbox{Re}\left\langle 
  \bar{z}_{\bm x} \cdot z_{{\bm x}+t\hat{\mu}}
\prod_{k=0}^{t-1} \lambda_{{\bm x}+k \hat{\mu},\hat\mu} 
\right\rangle\, .
\end{eqnarray}

\section{Numerical results} \label{sec4}

\subsection{Phase behavior for $N=4,10,15$}
\label{sec4.1}

\begin{figure}[tbp]
\includegraphics*[scale=\graphicscale,angle=-90]{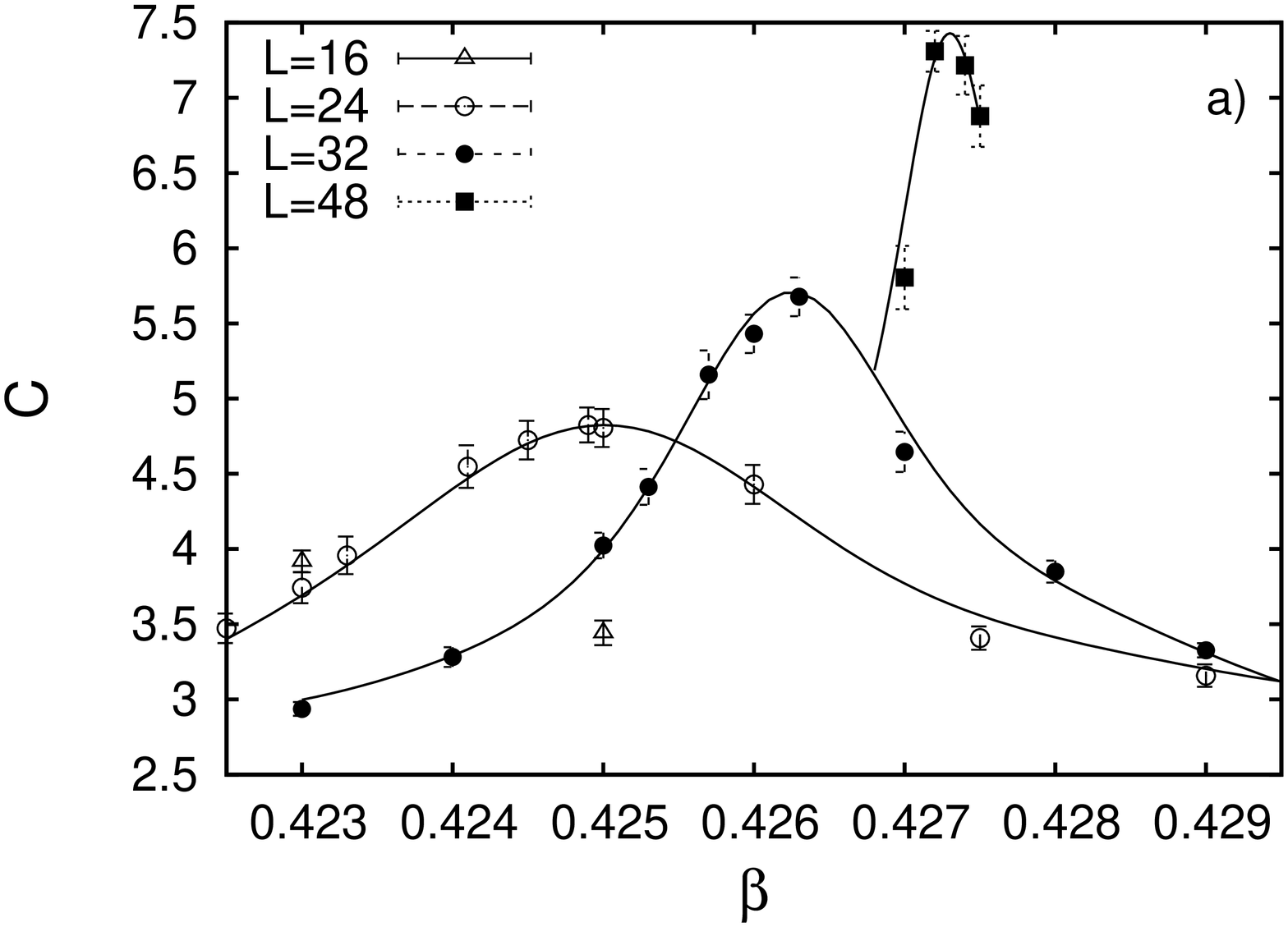}
\includegraphics*[scale=\graphicscale,angle=-90]{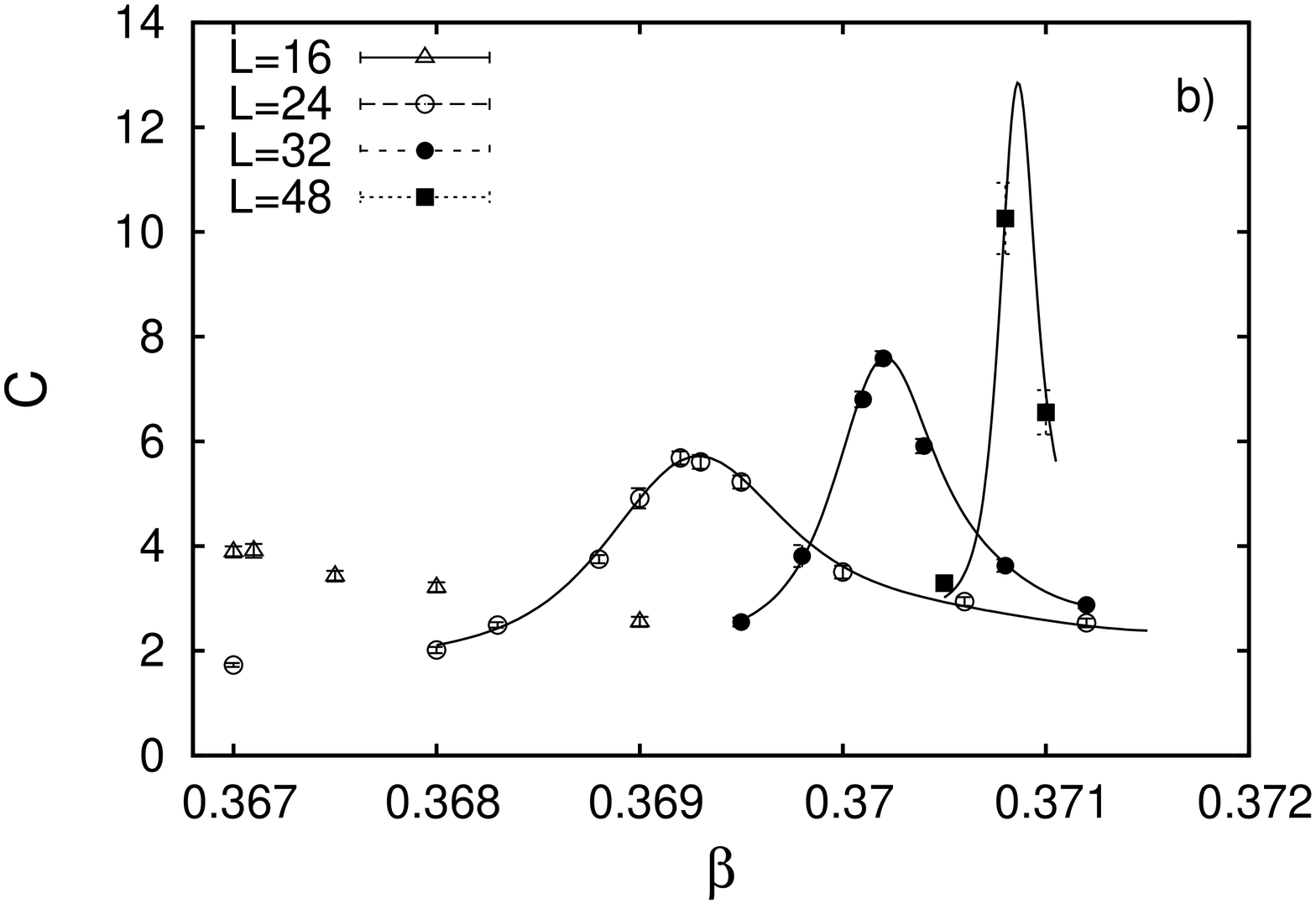}`
\includegraphics*[scale=\graphicscale,angle=-90]{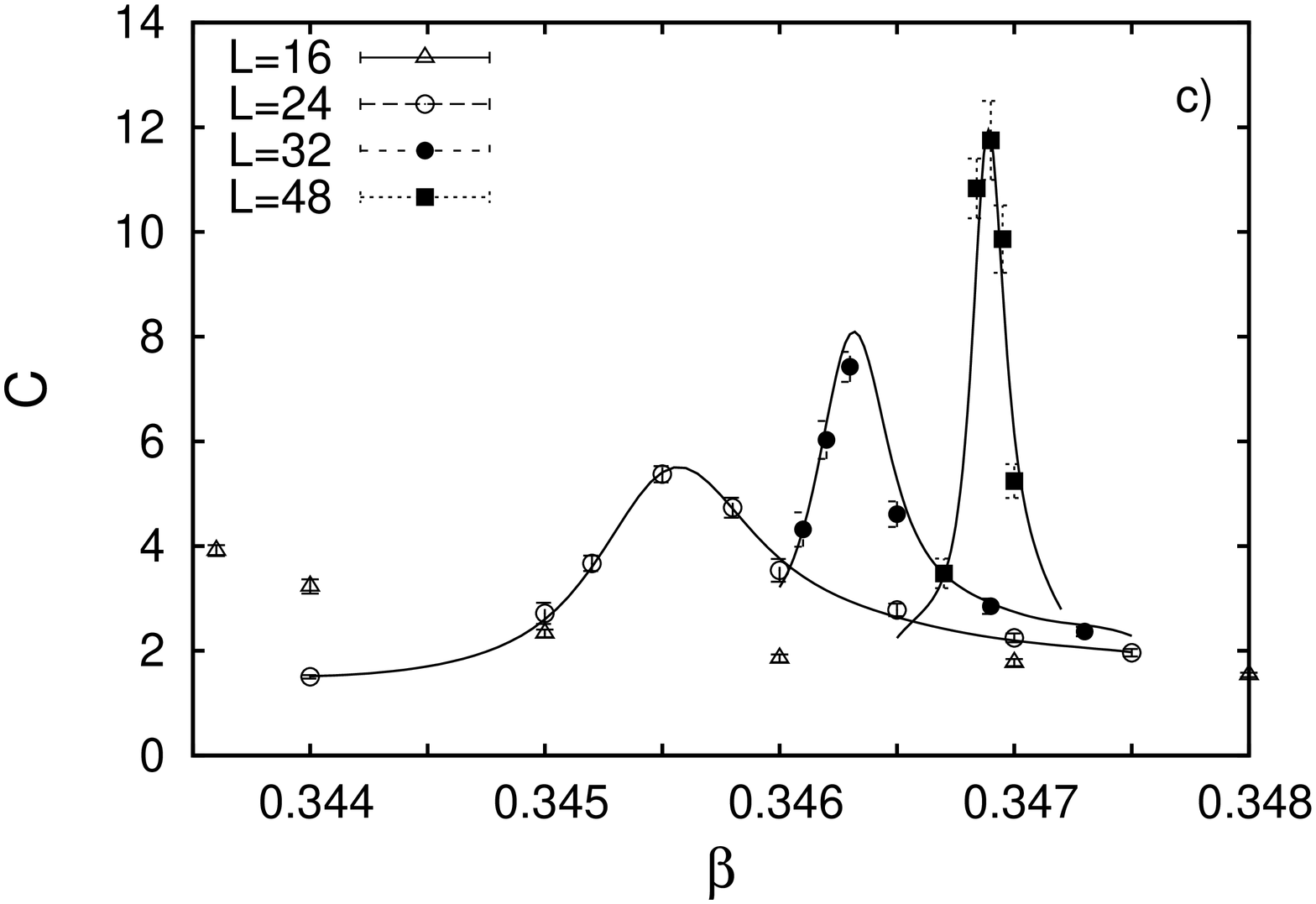}
\caption{Estimates of $C$ versus $\beta$ for the MFCP$^{N-1}$ model
  for $N=4$ (top, a), $N=10$ (middle, b), $N=15$ (bottom, c) and several
  lattice sizes $L$ up to $L=48$. The lines interpolating the data
  with $L=24, 32$, and $48$ are obtained using the multihistogram
  reweighting method \cite{FS-89}.  }
\label{Cv-N4N10N15}
\end{figure}

\begin{figure}[tbp]
\includegraphics*[scale=\graphicscale,angle=-90]{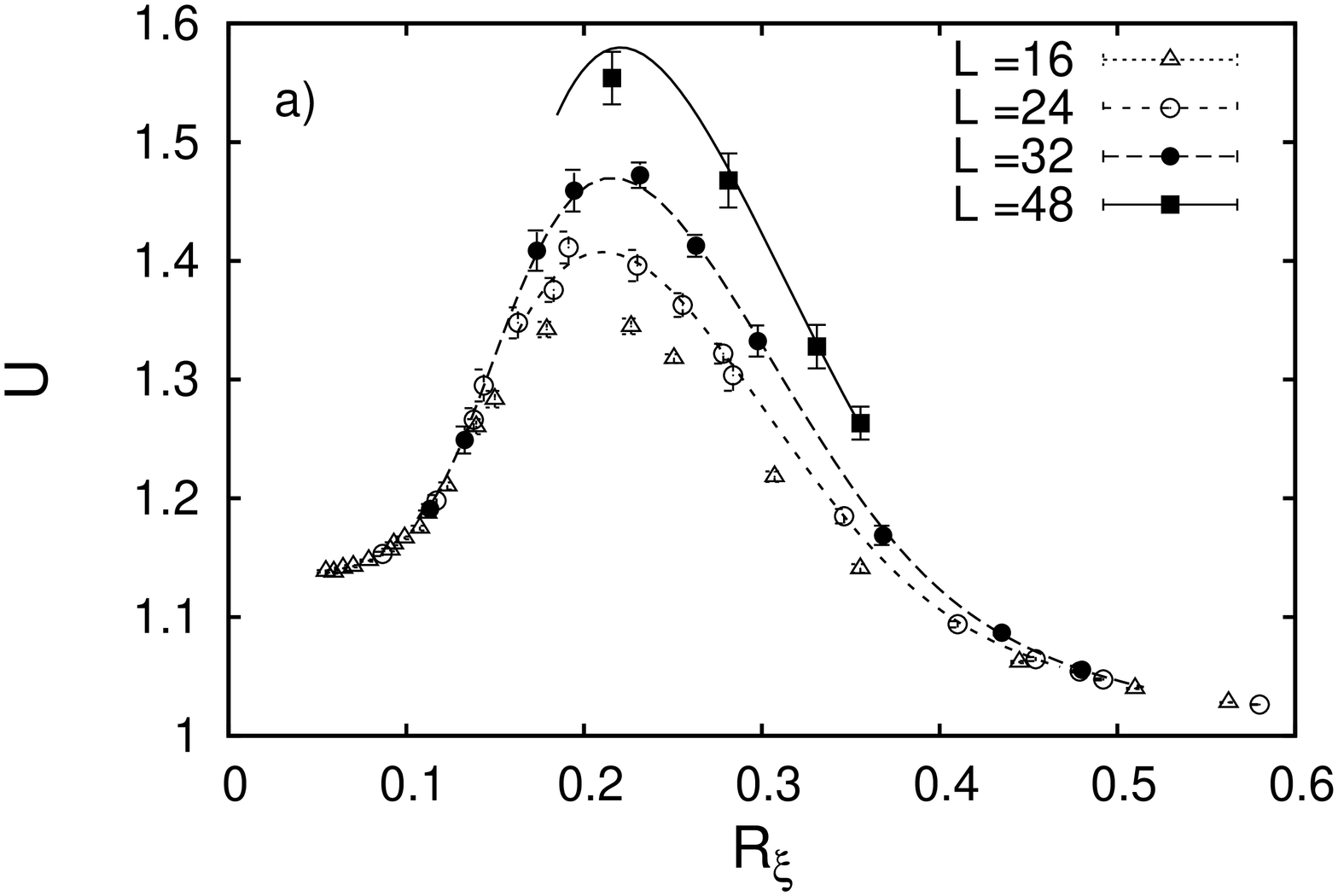}
\includegraphics*[scale=\graphicscale,angle=-90]{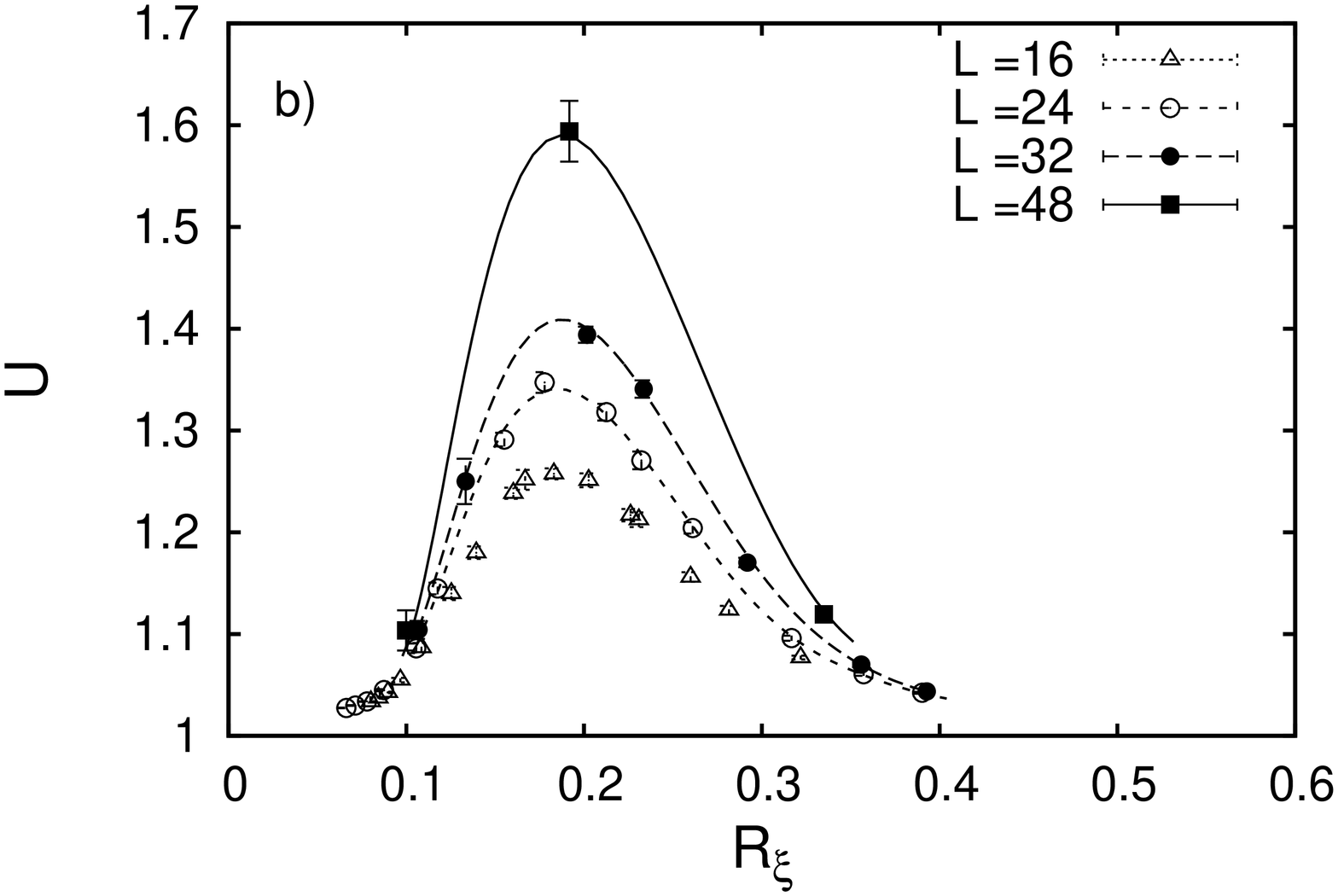}`
\includegraphics*[scale=\graphicscale,angle=-90]{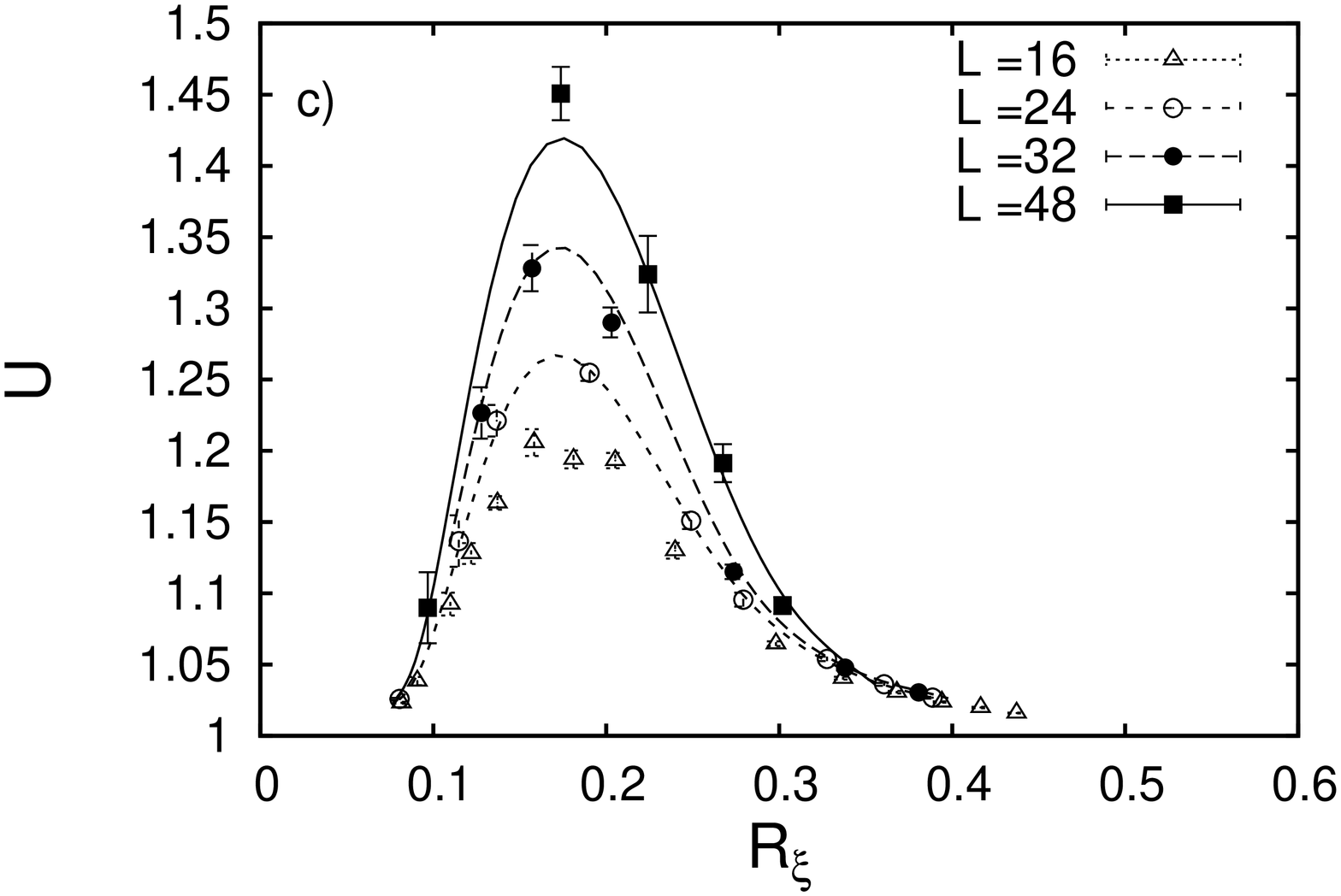}
\caption{Estimates of $U$ versus $R_\xi$ for the MFCP$^{N-1}$ model
  for $N=4$ (top, a), $N=10$ (middle, b), $N=15$ (bottom, c) and several
  lattice sizes $L$ up to $L=48$. The continuous lines interpolating the data
  with $L=24, 32$, and $48$ are obtained using the multihistogram
  reweighting method \cite{FS-89}. 
 }
\label{URxi-N4N10N15}
\end{figure}

We begin by discussing the behavior of the model for $N=4,10$, and
15. As we shall discuss, all results are consistent with a first-order
transition.  In Fig.~\ref{Cv-N4N10N15} we show the behavior of the
specific heat as a function of $\beta$. It shows clearly a maximum
that becomes larger and larger 
which increasing $L$, signaling the presence of a phase
transition. An estimate of the transition temperature can be obtained
by analyzing the Binder parameter $U$ as a function of
$\beta$. Irrespective of the nature of the transition ---it may be of
first order or continuous---the curves corresponding to different
sizes intersect at a temperature that converges to the transition
temperature as $L\to\infty$. We obtain $\beta_c = 
0.4285(5),0.3712(3),$ and $0.3472(3)$ for $N=4,10,15$, respectively.
These results are significantly lower that the transition values for
the model in which monopoles are allowed \cite{PV-19-CP,PV-20-largeN}:
$\beta_c = 0.5636(1),0.4253(5),0.381(1)$ for the same values of
$N$. This decrease of $\beta_c$ is expected, since the suppression of
monopoles gives rise to an effective ordering interaction, that makes
the high-temperature phase less stable.

From the data reported in Fig.~\ref{Cv-N4N10N15}, we can estimate the
maximum $C_{\rm max}(L)$ of the specific heat.  At a first-order
transition, it behaves as
\begin{equation}
C_{\rm max}(L) = {1\over 4} \Delta_h^2 V \left[ 1 + O(V^{-1})\right],
\label{Cmax-first}
\end{equation}
where $V=L^d$  is the $d$-dimensional volume ($d=3$) and $\Delta_h$ is the latent
heat.  At a continuous transition, instead, we have
\begin{equation}
C_{\rm max}(L) = a L^{\alpha/\nu} + C_{\rm reg},
\end{equation}
where the constant term $C_{\rm reg}$ is due to the analytic
background. It is the dominant contribution if $\alpha < 0$.  If we
fit $C_{\rm max}(L)$ with a simple power behavior $a L^\delta$, we
obtain $\delta = 0.7(2)$, 1.3(2), 1.1(3) for $N=4,10,15$.  This
behavior is quite different from that expected at a first-order
transition ($\delta = d = 3$). If we assume that the transition is
continuous, we should have $\delta = \alpha/\nu$, which would give 
$\nu = 0.54(3)$,
0.47(3), 0.49(4)  (we
use the hyperscaling relation $2 - \alpha = d \nu$) 
for $N=4,10,15$, respectively.

The large difference between the estimates of $\delta$ and the
first-order prediction $\delta = 3$ might be taken, a priori, as an
indication that the transition is continuous. However, experience with
similar models that undergo weak first-order transitions indicates
that in many cases the analysis of the specific heat is not
conclusive. The behavior (\ref{Cmax-first}) may set in at values of
$L$ that are much larger than those at which simulations can be actually
performed.  In the case of weak first-order transitions, a more useful
quantity is the Binder parameter $U$. At a first-order transition, the
maximum $U_{\rm max}(L)$ of $U$ for each size $L$ behaves
as~\cite{CLB-86,VRSB-93}
\begin{equation}\label{Ufirst}
U_{\rm max}(L)= c \,V \left[1 + O(V^{-1})\right]\,.
\end{equation}
On the other hand, 
$U$ is bounded as $L\to \infty$ at a continuous phase transition.
Indeed, at such transitions, in the FSS limit, any
renormalization-group invariant quantity $R$ scales as
\begin{equation}
R(\beta,L) = f_R(X) + O(L^{-\omega}) ,\quad X = (\beta-\beta_c) L^{1/\nu} ,
\label{rsca}
\end{equation}
where $f_R(X)$ is a regular function, which is universal apart from a
trivial rescaling of its argument, and $\omega$ is a correction to
scaling exponent.  Therefore, $U$ has a qualitatively different
scaling behavior for first-order or continuous transitions. In practice,
a first-order transition can be simply identified by verifying that
$U_{\rm max}(L)$ increases with $L$, without the need of explicitly
observing the linear behavior in the volume.

In the case of weak first-order transitions, the nature of the
transition can also be understood from the combined analysis of $U$
and $R_{\xi}$ \cite{PV-19-CP}. At a continuous transition, in the FSS
limit the Binder parameter $U$ (more generally, any
renormalization-group invariant quantity) can be expressed in terms of
$R_\xi$ as
\begin{equation}
U(\beta,L) = F_R(R_\xi) + O(L^{-\omega}),
\label{r12sca}
\end{equation}
where $F_R(x)$ is universal. This scaling relation does not hold at
first-order transitions, because of the divergence of $U$ for
$L\to\infty$.  Therefore, the order of the transition can be
understood from plots of $U$ versus $R_\xi$.  The absence of a data
collapse is an early indication of the first-order nature of the
transition, as already advocated in Ref.~\cite{PV-19-CP}.

To understand the order of the transition, in Fig.~\ref{URxi-N4N10N15}
we report the Binder parameter as a function of $R_\xi$.  The observed
behavior is not consistent with a continuous transition.  Data do not
scale and moreover, the Binder parameter has a maximum that increases
with the size $L$, a behavior that can only be observed at first-order
transitions.

\begin{figure}[tbp]
\includegraphics*[scale=\graphicscale,angle=-90]{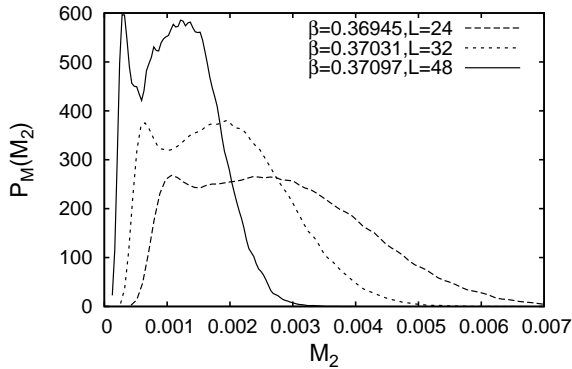}
\caption{Distribution $P_M(M_2)$ for the MFCP$^9$ model for
  different values of $L$.  For each value of $L$ we report the
  distribution for the value of $\beta$ at which $P_M(M_2)$ has two
  peaks of approximately the same height.  The distributions are
  obtained using the multihistogram reweighting method \cite{FS-89}.
}
\label{PM2-N10}
\end{figure}

\begin{figure}[tbp]
\includegraphics*[scale=\graphicscale,angle=-90]{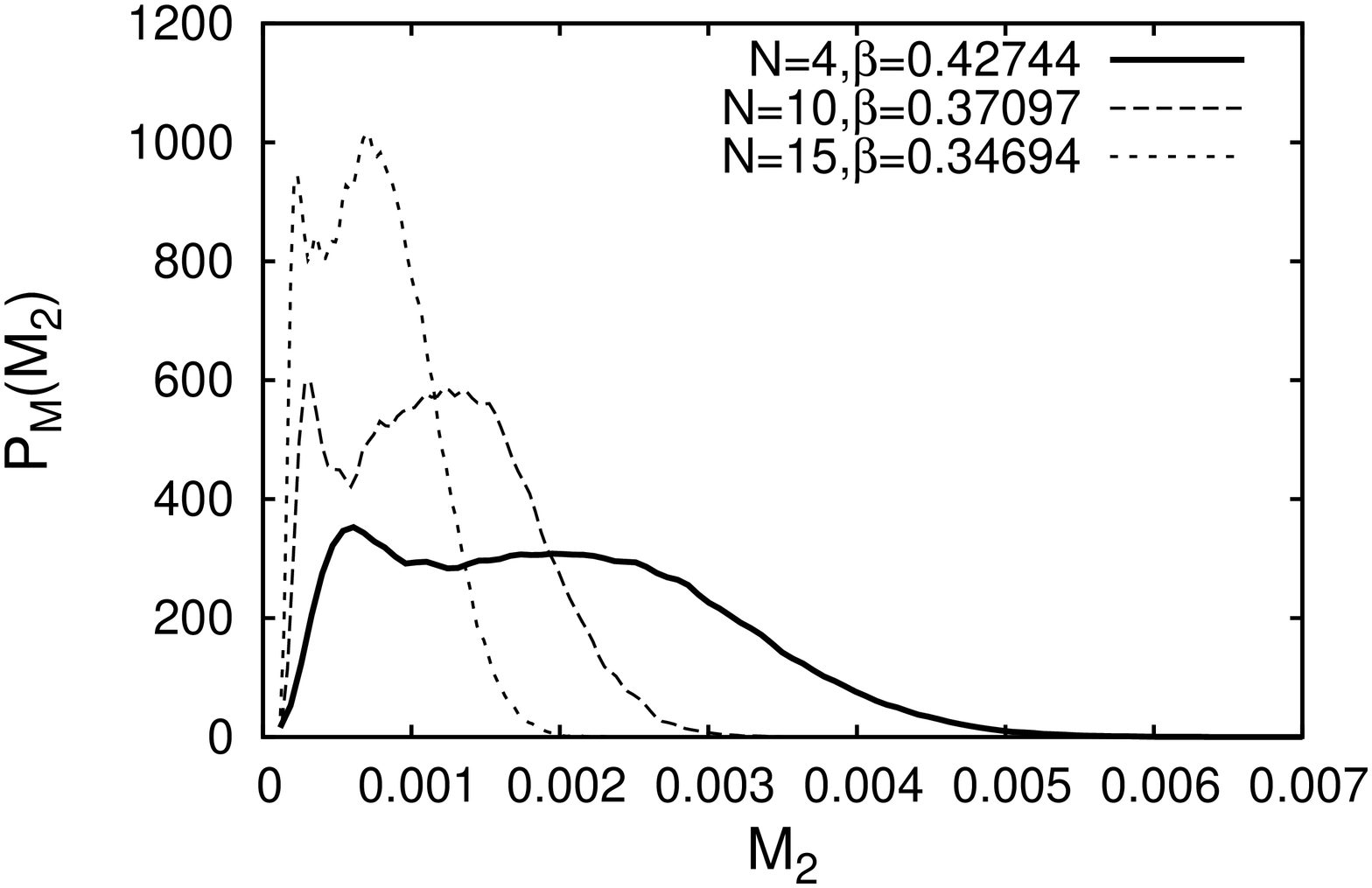}
\caption{Distribution $P_M(M_2)$ for the MFCP$^{N-1}$ model for $L=48$
  and different values of $N$.  For each value of $N$ we report the
  distribution for the value of $\beta$ at which $P_M(M_2)$ has two
  peaks of approximately the same height.  The distributions are
  obtained using the multihistogram reweighting method \cite{FS-89}.
}
\label{PM2-L48}
\end{figure}

\begin{figure}[tbp]
\includegraphics*[scale=\graphicscale,angle=-90]{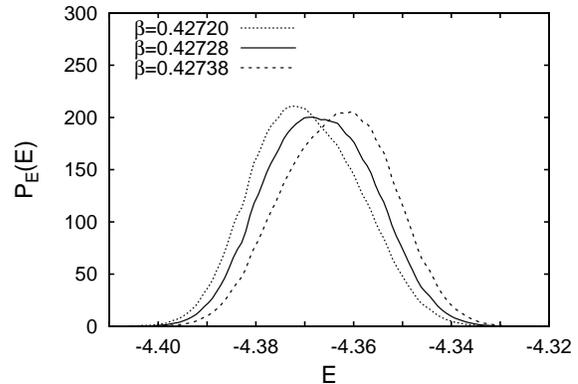}
\caption{Distribution $P_E(E)$ for the MFCP$^3$ model for $L=48$
  and different values of $\beta$.
  The distributions are obtained using the multihistogram 
  reweighting method \cite{FS-89}.
}
\label{PE-N4}
\end{figure}

To further confirm the discontinuous nature of the transition we have
studied the distributions of the order parameter and of the energy:
\begin{eqnarray}
&&P_E(E) = \langle \delta[E - H/(NV)]\rangle, \\
&&P_M(M_2) = \langle \delta(M_2 -\mu_2) \rangle,
\nonumber
\end{eqnarray}
where $\mu_2$ is defined in Eq.~(\ref{binderdef}).  In
Fig.~\ref{PM2-N10} we show $P_M(M_2)$ for $N=10$ and several values of
$L$. For each size, we consider the value of $\beta$ at which the
distribution shows two peaks of approximately the same height $P_{\rm
  max}$.  As expected for a first-order transition, if $P_{\rm min}$
is the minimum of the distribution between the two maxima, we observe
that the ratio $P_{\rm max}/P_{\rm min}$ increases with $L$. This
increase is not consistent with a continuous transition. Indeed, at
such transitions the distribution $P_M(M_2)$ may have two peaks ---
this is the case for the 3D Ising model \cite{TB-00}.  However, in
the Ising  case the ratio $P_{\rm max}/P_{\rm min}$ is constant in the
large-$L$ limit.  It is worth noting that the transition is very
weak. The ratio $P_{\rm max}/P_{\rm min}$ is only slightly larger than
one (the dip is barely significant if we take statistical errors into
account) for $L=24$, and is approximately equal to 1.2 and 1.4 for
$L=32$ and 48.  This implies that there is still a significant overlap
between the two phases, which explains the  strong size dependence
of the distribution.  It is important to note that the distribution
has two peaks only in a very tiny $\beta$ interval. For $L=48$, they
are observed only when $\beta$ belongs to the interval
$[0.3709,0.3710]$. Therefore, we made extensive use of the
multihistogram method of Ref.~\cite{FS-89}, which allowed us to
compute the distributions on a very fine grid of $\beta$ values.

For both $N=4$ and $N=15$, the transition is weaker than for $N=10$.
We observe two peaks only for $L=48$ in the first case and for
$N=32,48$ in the second one. This is evident from the results reported
in Fig.~\ref{PM2-L48}, where we show results for different values of
$N$ and $L=48$. For $N=4$ two peaks are barely visible, while for
$N=15$ we have $P_{\rm max}/P_{\rm min}\approx 1.15$. As a second
remark, note also that the distributions become more narrow as $N$
increases, indicating that the spontaneous magnetization decreases as
$N$ becomes large.  We have also studied the distributions for the
energy. For $N=4$ a double-peak structure is not observed even for
$L=48$, although there is some evidence of two-phase behavior, see
Fig.~\ref{PE-N4}. A double-peak structure is instead observed for both
$N=10$ and $N=15$.

\subsection{Results for $N=2$} \label{sec4.2}

\begin{figure}[tbp]
\includegraphics*[scale=\graphicscale,angle=-90]{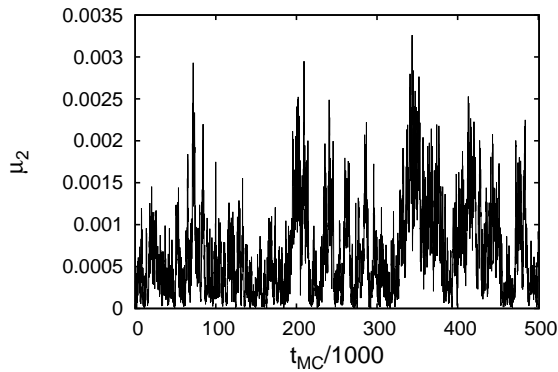}
\caption{Time evolution of $\mu_2$ for $\beta = 0.4598$, $L=80$,
  $N=2$.  Time is measured in lattice sweeps. We report the results
  for a time interval of $5\cdot 10^5$ sweeps.}
\label{MCevolution-N2}
\end{figure}

\begin{figure}[tbp]
\includegraphics*[scale=\graphicscale,angle=-90]{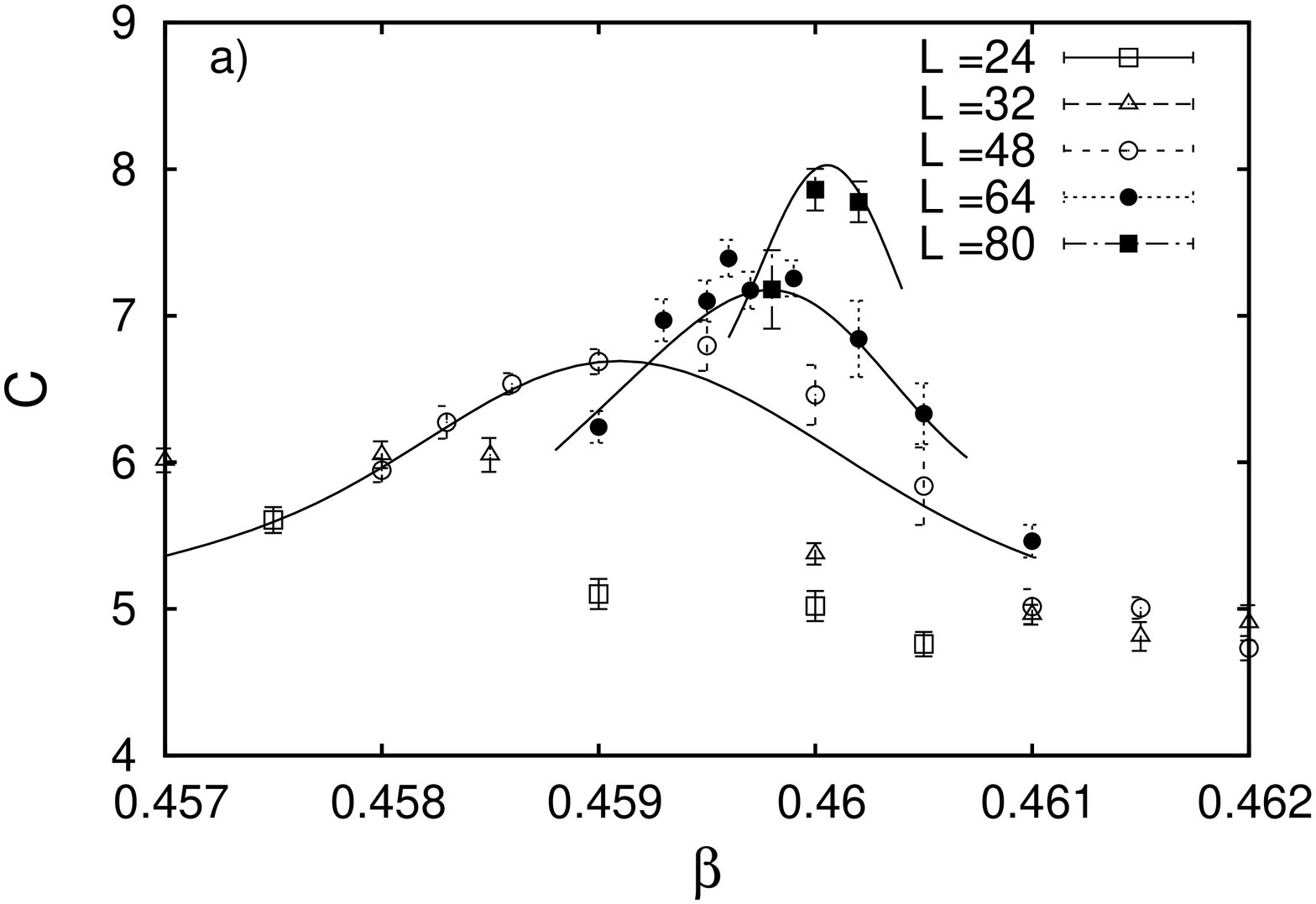}
\includegraphics*[scale=\graphicscale,angle=-90]{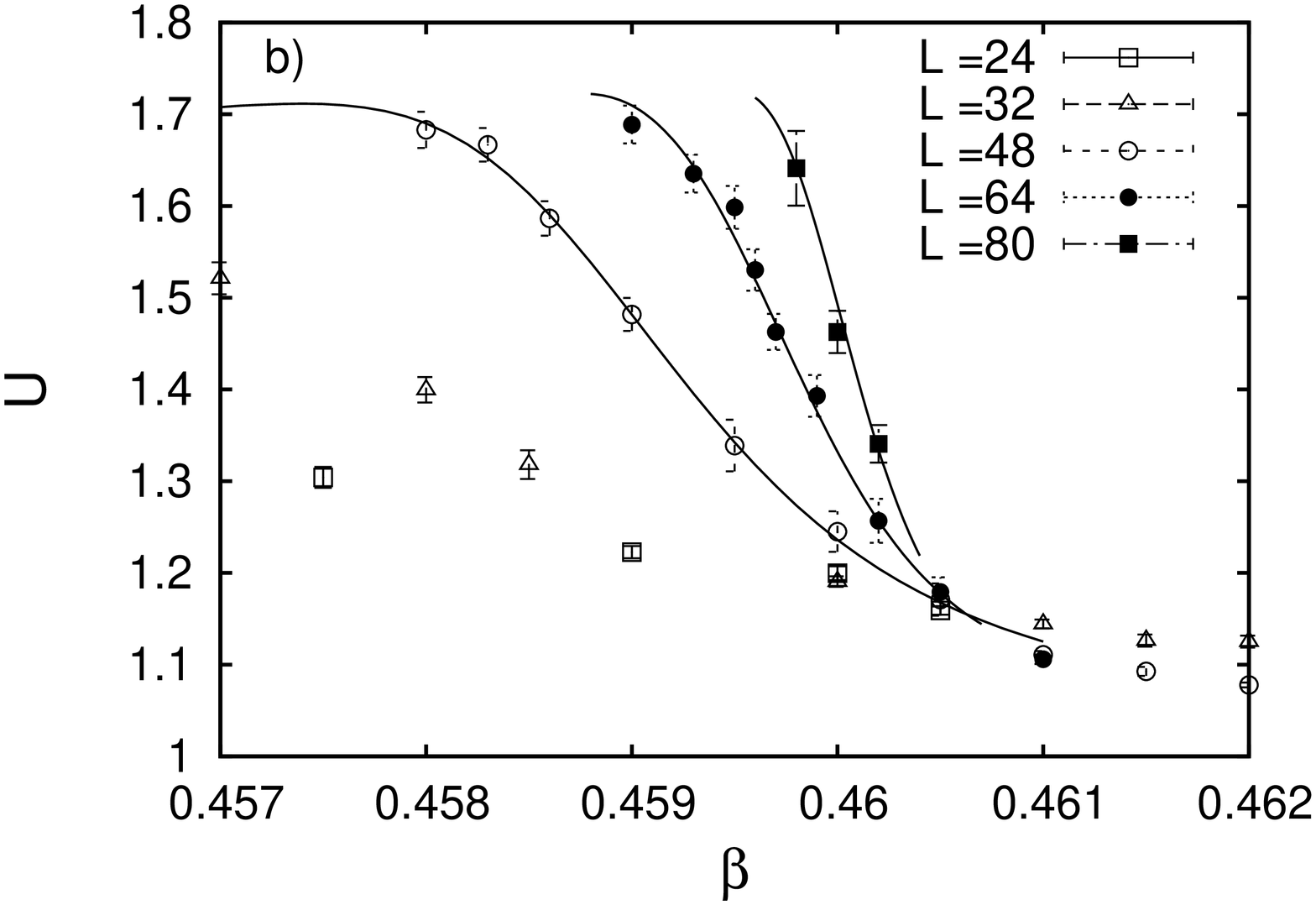}`
\caption{Plot of the specific heat $C$ (top, a) and of the Binder
  parameter $U$ (bottom, b) as a function of $\beta$ in the transition
  region.  Results for several values of $L$ up to $L=80$ for $N=2$.
  The curves (continuous lines)
  interpolating the data with $L=48$, 64, and 80 are
  obtained using the multihistogram reweighting method \cite{FS-89}.
}
\label{CVU-beta-N2}
\end{figure}

Let us now discuss the results for $N=2$. In this case we have not
been able to draw any definite conclusion on the order of the
transition. We have performed extensive simulations on lattices of
size up to $L=80$. Each data point consists in $N_{\rm sw}$
lattice sweeps, with $N_{\rm sw}$ varying between $10^6$ and $5\cdot
10^6$. In spite of the large number of iterations, the statistics is
not large, especially for $L\ge 48$, since autocorrelation times are
huge. For $L=64$ and 80, the integrated autocorrelation time
associated with $\mu_2$ [see Eq.~(\ref{binderdef})] is of order 3000
and 5000 iterations, respectively, in the transition region, so that
the number of independent configurations varies between 500 and 1000
for these two values of $L$.  The presence of strong autocorrelations
can be easily understood by looking at the time dependence of $\mu_2$
reported in Fig.~\ref{MCevolution-N2}.  Typical configurations are not
magnetized---$\mu_2$ is very small---but, at intervals of the order of
$10^3$-$10^4$ iterations, a fluctuation occurs towards configurations
of larger magnetization. In order to improve the quality of the
results, we have extensively used the multihistogram method of
Ref.~\cite{FS-89}, combining all runs corresponding to the same size
$L$.

In Fig.~\ref{CVU-beta-N2} we report the specific heat $C$ and the
Binder parameter $U$ as a function of $\beta$. The parameter $U$ shows
an intersection for $\beta\approx 0.4605$, indicating the presence of
a phase transition.  In the same $\beta$ region the specific heat has
a peak that increases with the size $L$.  For each value of $L$ we
have determined $C_{\rm max}(L)$. A fit of the results for $L\ge 48$
to $a L^\delta$ gives $\delta = 0.35(8)$. We have also performed a fit
including an analytic correction, fitting $\ln C_{\rm max}(L)$ to 
$\delta \ln L + a + b L^{-\delta}$. Using the results for $L\ge 32$ we obtain
$\delta = 0.7(2)$. 
The exponent $\delta$ is
quite different from what one would expect for a first-order
transition, $\delta = 3$. If the transition is continuous, $\delta$
should be identified with $\alpha/\nu$. Using the hyperscaling
relation $2-\alpha = 3\nu$, we would then predict $\nu =
0.60(2)$ and $0.54(3)$, using the two results for $\delta$. 
Note that the results for the specific heat exclude a
critical transition in the O(3) universality class, since $\alpha < 0$
for the latter model \cite{PV-02}.

\begin{figure}[tbp]
\includegraphics*[scale=\graphicscale,angle=-90]{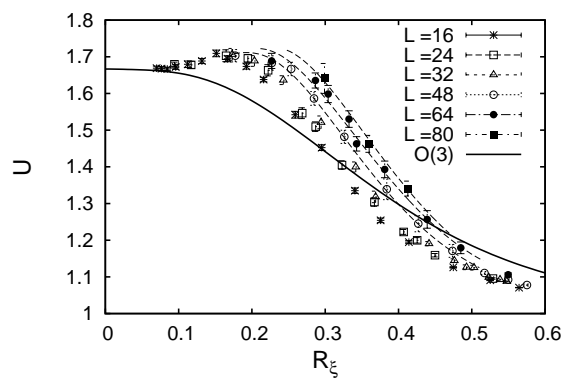}
\caption{Plot of the Binder parameter $U$ versus $R_\xi$, for several
  values of $L$ up to $L=80$ for $N=2$.  The curves (dashed lines)
  interpolating the
  data with $L=48$, 64, and 80 ($L$ increases moving rightward) are
  obtained using the multihistogram reweighting method \cite{FS-89}.
  The continuous thicker line is the universal curve for the O(3)
  universality class \cite{footnoteO3}.  }
\label{URxi-N2}
\end{figure}

\begin{figure}[tbp]
\includegraphics*[scale=\graphicscale,angle=-90]{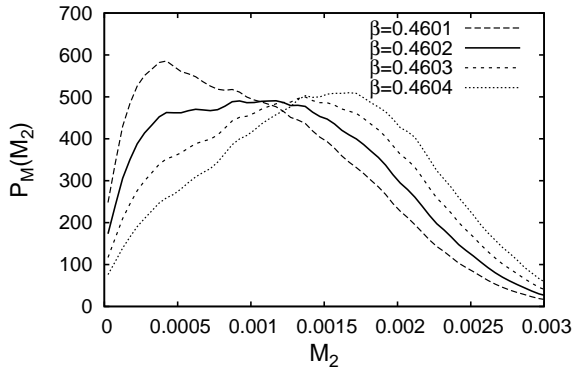}
\caption{Distribution $P_M(M_2)$ for the MFCP$^1$ model for $L=80$ and
  three different values of $\beta$.  The distributions are obtained
  using the multihistogram reweighting method \cite{FS-89}.  }
\label{PM2-N2}
\end{figure}

To understand the nature of the transition, in Fig.~\ref{URxi-N2} we
report $U$ versus $R_\xi$. In this case, the plot does not allow us to
draw any definite conclusion. On one side, the data do not scale: At
fixed $R_\xi$, the estimates of $U$ are systematically increasing with
$L$ for $0.2\le R_\xi\lesssim 0.5$. This would favor a
first-order transition. On the other hand, the estimates of $U$ do not
show a maximum that increases with $L$. This behavior is 
usually taken as an indication for 
a continuous transition, although the recent results of Ref.~\cite{SN-19}
show that it is possible to have a discontinuous transition even when 
the Binder parameter does not show a peak 
for lattice sizes that are usually
considered quite large (they perform simulations up to $L=256$).
Whatever the interpretation be, O(3) behavior
is clearly excluded, as already noted from the analysis of the
specific heat.

We have also computed the distributions of $\mu_2$. We do not observe
any double-peak structure. However, as $\beta$ varies, the
distributions change as expected for a first-order transition, see
Fig.~\ref{PM2-N2}. Indeed, for $\beta = 0.4901$, $P_{M}(M_2)$ has a
peak for $M_2 = 0.0004$, for $\beta = 0.4902$ the curve flattens, and
then is starts showing a new distinct maximum at $M_2\approx 0.002$ as
$\beta$ increases. This behavior is consistent with what is observed
in Fig.~\ref{MCevolution-N2}. The large fluctuations can be
interpreted as the typical seesaw behavior observed in the presence of
two distinct coexisting phases. The system moves between the
unmagnetized phase ($\mu_2 \sim 10^{-4}$) and a magnetized phase with
$\mu_2 \approx 0.002$.

\begin{table}[t]
\caption{Results of the fits to Eq.~(\ref{rsca}) for different values
  of $L_{\rm min}$, the minimum size of the data included in the fit.
  For the function $f(x)$ we take a twelfth-order polynomial for
  $L_{\rm min} =24$ and 32, and a sixth-order polynomial for $L_{\rm
    min} = 48$. Here $\chi^2$ is the sum of the fit square residuals
  and DOF is the number of degrees of freedom.  }
\label{fits-nubetac}
\begin{tabular}{lcccc}
\hline\hline
   & $L_{\rm min}$  & $\chi^2$/DOF & $\nu$ & $\beta_c$ \\
\hline
$U_4$  & 24  & 39/33 &  0.547(5)  & 0.46057(3) \\
       & 32  & 22/22 &  0.516(17) & 0.46054(4) \\
       & 48  &  8/14 &  0.56(6)   & 0.46066(10) \\
$R_\xi$& 24  &160/33 &  0.594(4)  & 0.46035(1) \\
       & 32  & 58/22 &  0.527(9)  & 0.46030(2) \\
       & 48  &  6/14 &  0.53(3)   & 0.46040(5) \\
\hline\hline
\end{tabular}
\end{table}

\begin{figure}[tbp]
\includegraphics*[scale=\graphicscale,angle=-90]{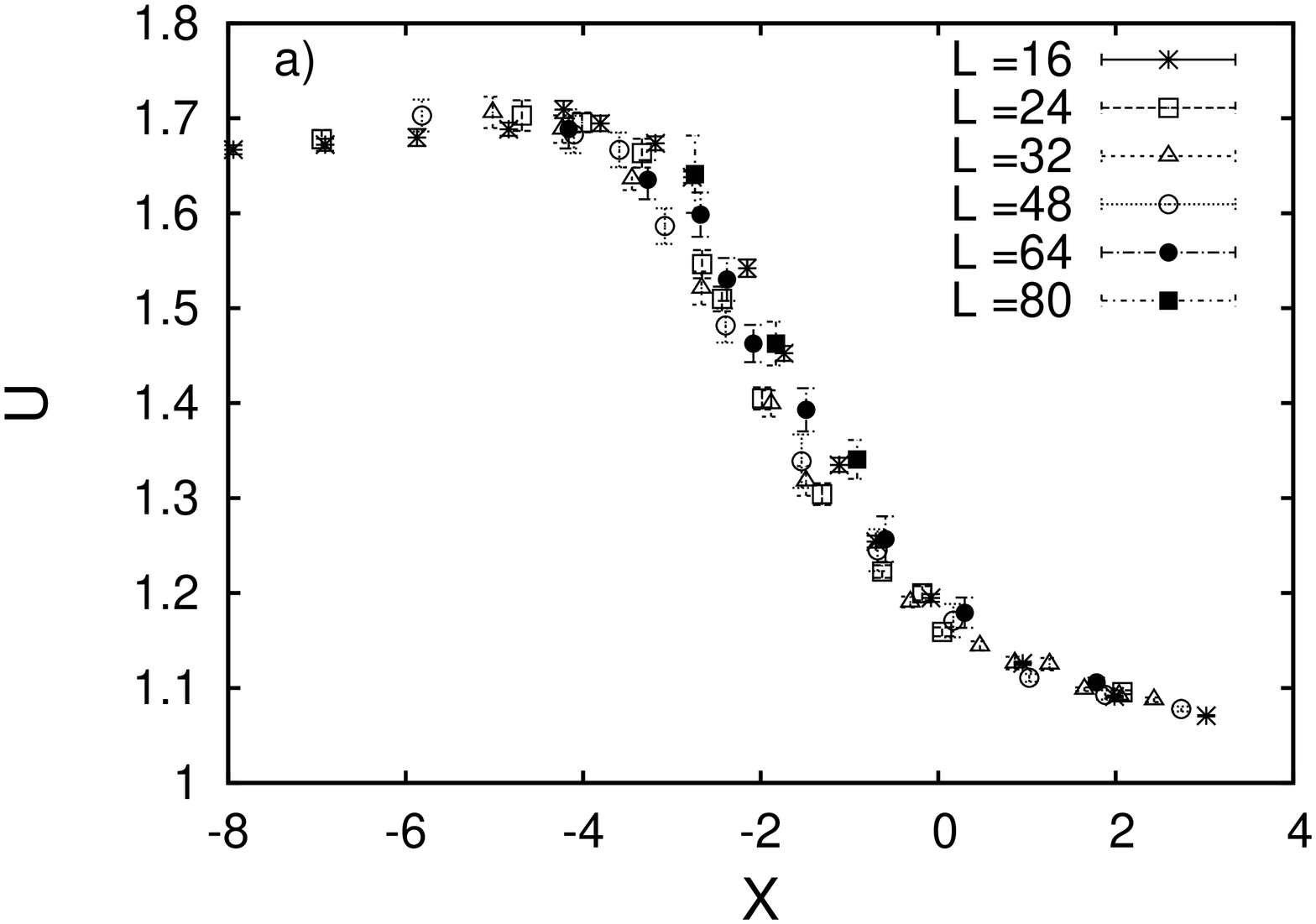}
\includegraphics*[scale=\graphicscale,angle=-90]{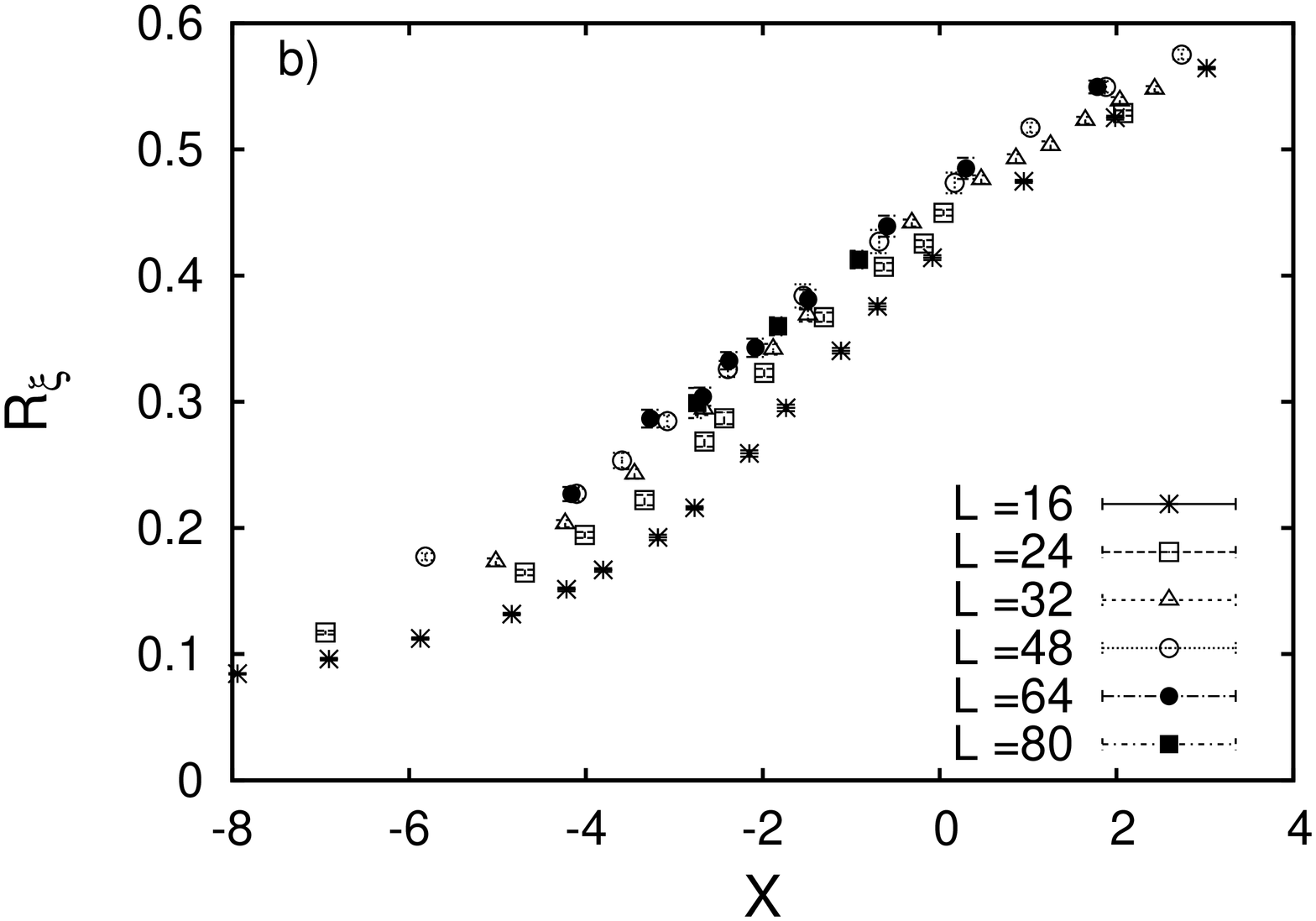}
\caption{Plot of $U$ (top, a) and of $R_{\xi}$ (bottom, b) as a function of 
  $X = (\beta - \beta_c) L^{1/\nu}$, using $\beta_c = 0.4605$ and $\nu = 0.52$,
  the estimates (\ref{estimates-N2}).
}
\label{scalingplot-N2}
\end{figure}

To conclude the analysis of the available data, we may assume that the
transition is continuous and determine the critical exponents. First,
we determine $\nu$ and the transition value $\beta_c$ fitting the data
to Eq.~(\ref{rsca}). The function $f_R(x)$ is approximated by a
polynomial.  The results of the fits are reported in
Table~\ref{fits-nubetac} as a function of $L_{\rm min}$, the minimum
size of the data included in the fit. We observe a significant drop of
the estimate of $\nu$ as $L_{\rm min}$ increases from 24 to 32. This
is due to the large scaling corrections we have already observed when
considering $U$ versus $R_\xi$, Moreover, the estimates of $\beta_c$
obtained by using $R_\xi$ and $U$ are not consistent within errors.
If we average the results of the two analyses, we would estimate
\begin{equation}
\nu = 0.52(2), \qquad \beta_c = 0.4605(3).
\label{estimates-N2}
\end{equation}
In Fig.~\ref{scalingplot-N2} we report the corresponding scaling
plots.  As expected, the quality of the scaling is poor: large
deviations are present. In any case, note that the estimate of $\nu$ is 
consistent with that obtained from the specific heat, $\nu = 0.54(3)$,
obtained including the analytic corrections.

\begin{figure}[tbp]
\includegraphics*[scale=\graphicscale,angle=-90]{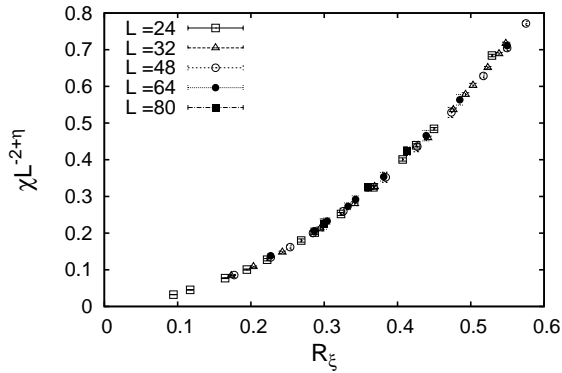}
\caption{Plot of $\chi/L^{2-\eta}$  as a function of 
  $R_\xi$, using $\eta = 0.335$.
}
\label{scalingplotchi-N2}
\end{figure}

Finally, we determine the exponent $\eta$ associated with the
susceptibility $\chi$ defined in Eq.~(\ref{chisusc}).  This quantity
scales as
\begin{equation}
\chi(\beta,L) \sim L^{2-\eta} \left[ f_\chi(X) + O(L^{-\omega})\right],
\label{chisca}
\end{equation}
or, equivalently, as
\begin{equation}
\chi(\beta,L) \sim L^{2-\eta} \left[ F_\chi(R_\xi) + O(L^{-\omega})\right].
\label{chisca2}
\end{equation}
We therefore fit the data to $\ln \chi = (2 - \eta) \log L +
\hat{f}_\chi(R_\xi)$, where we approximate the function
$\hat{f}_\chi(x)$ with a polynomial in $x$.  To estimate the role of
the scaling corrections we include in the fit only the data
corresponding to sizes $L\ge L_{\rm min}$. We obtain $\eta = 0.352(7)$
and 0.335(10) for $L_{\rm min} = 24$ and 32, respectively. In this
case, scaling corrections appear to be small ($\chi^2$/DOF is approximately
0.99 for $L_{\rm min} = 24$ and 0.56 for $L_{\rm min} = 32$; DOF 
is the number of degrees of freedom of the fit), as is also evident from
the scaling plot, Fig.~\ref{scalingplotchi-N2}. Conservatively, 
we will take $\eta = 0.335(10)$ as our final estimate.

In conclusion the results for $N=2$ may be still interpreted in terms
of two different scenarios. A first possibility is that the transition
is of first order.  This would explain the poor scaling we observe
when we plot $U$ versus $R_\xi$, the inconsistencies in the results of
$\nu$ and $\beta_c$ obtained in the analysis of $U$ and $R_\xi$, 
and the shape of the distribution of the order
parameter, see Fig.~\ref{PM2-N2}. However, the absence of a divergence
in the behavior of the Binder parameter does not allow us to exclude
that the transition is continuous and that the observed
inconsistencies are simply due to scaling corrections that are
particularly large in this model. 
A continuous transition is also
supported by the behavior of the susceptibility that shows a good
scaling, which allows us to obtain an apparently accurate estimate of the
exponent $\eta$. If the transition is continuous, it does not belong to 
the Heisenberg universality class: O(3) behavior is clearly excluded by the
data.

\subsection{Results for $N=25$} \label{sec4.3}

\begin{figure}[tbp]
\includegraphics*[scale=\graphicscale,angle=-90]{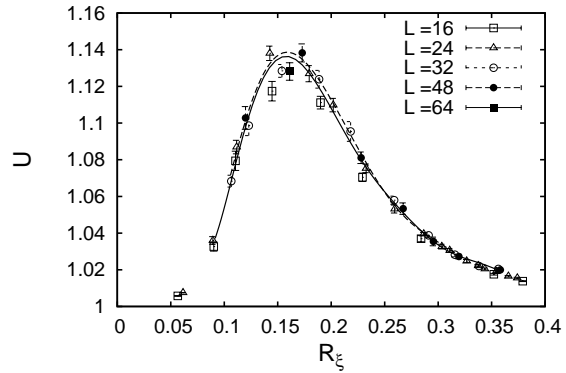}
\caption{Plot of the Binder parameter $U$ versus $R_\xi$, for several
  values of $L$ up to $L=64$ for $N=25$.  The curves
  interpolating the
  data with $L=32$ (dashed line) and 48 (continuous line) are obtained
  using the multihistogram reweighting method \cite{FS-89}.  }
\label{URxi-N25}
\end{figure}

\begin{table}[t]
\caption{Results of the fits to Eq.~(\ref{rsca}), as a function of 
$L_{\rm min}$. For the function $f_R(x)$ 
we take a twelfth-order polynomial. Results for the MFCP$^{24}$ model.
}
\label{fits-nubetac-N25}
\begin{tabular}{lcccc}
\hline\hline
   & $L_{\rm min}$  & $\chi^2$/DOF & $\nu$ & $\beta_c$ \\
\hline
$R_\xi$ & 16  & 89/31 &  0.579(2)  & 0.319943(3) \\
       & 24  & 39/23 &  0.597(4)  & 0.319963(5) \\
$U_4$  & 16  &112/31 &  0.567(5)  & 0.319938(10) \\
       & 24  & 48/23 &  0.593(9)  & 0.319969(13) \\
\hline\hline
\end{tabular}
\end{table}

We finally present our results for $N=25$. We have performed
simulations on lattices of size $16\le L \le 64$. Autocorrelations are
very large (of the order of $10^3$ for $L = 64$), so that simulations
of larger lattices are unfeasible. Note that most of the data
correspond to $L\le 48$.  For $L=64$ we have a single data point.  As
we shall discuss, all results are consistent with a continuous
transition.

We first analyze the behavior of the Binder parameter and of the
specific heat as a function of $\beta$. The specific heat shows a very
clear maximum that increases with $L$ and the Binder parameter curves
at fixed $L$ have a crossing point for $\beta \approx 0.320$, which
allows us to identify the transition region. To determine the order of
the transition, we consider the plot of $U$ versus $R_\xi$, see
Fig.~\ref{URxi-N25}. It is quite evident that all results
approximately fall onto a single curve with small scaling
corrections. This is confirmed by the curves obtained by using the
multihistogram reweighting method of Ref.~\cite{FS-89}: the curves
corresponding to $L=32$ and 48 cannot be distinguished on the 
scale of figure except at the
peak. Note that the curves apparently indicate that $U_{\rm max}(L)$
decreases as $L$ is increased, which is the opposite behavior of that
expected at first-order transitions. The downward trend at the peak is
also confirmed by the result obtained for $L=64$: the estimate of $U$
is lower than the $L=48$ curve, see Fig.~\ref{URxi-N25}. We can thus
exclude that the transition is of first order.

\begin{figure}[tbp]
\includegraphics*[scale=\graphicscale,angle=-90]{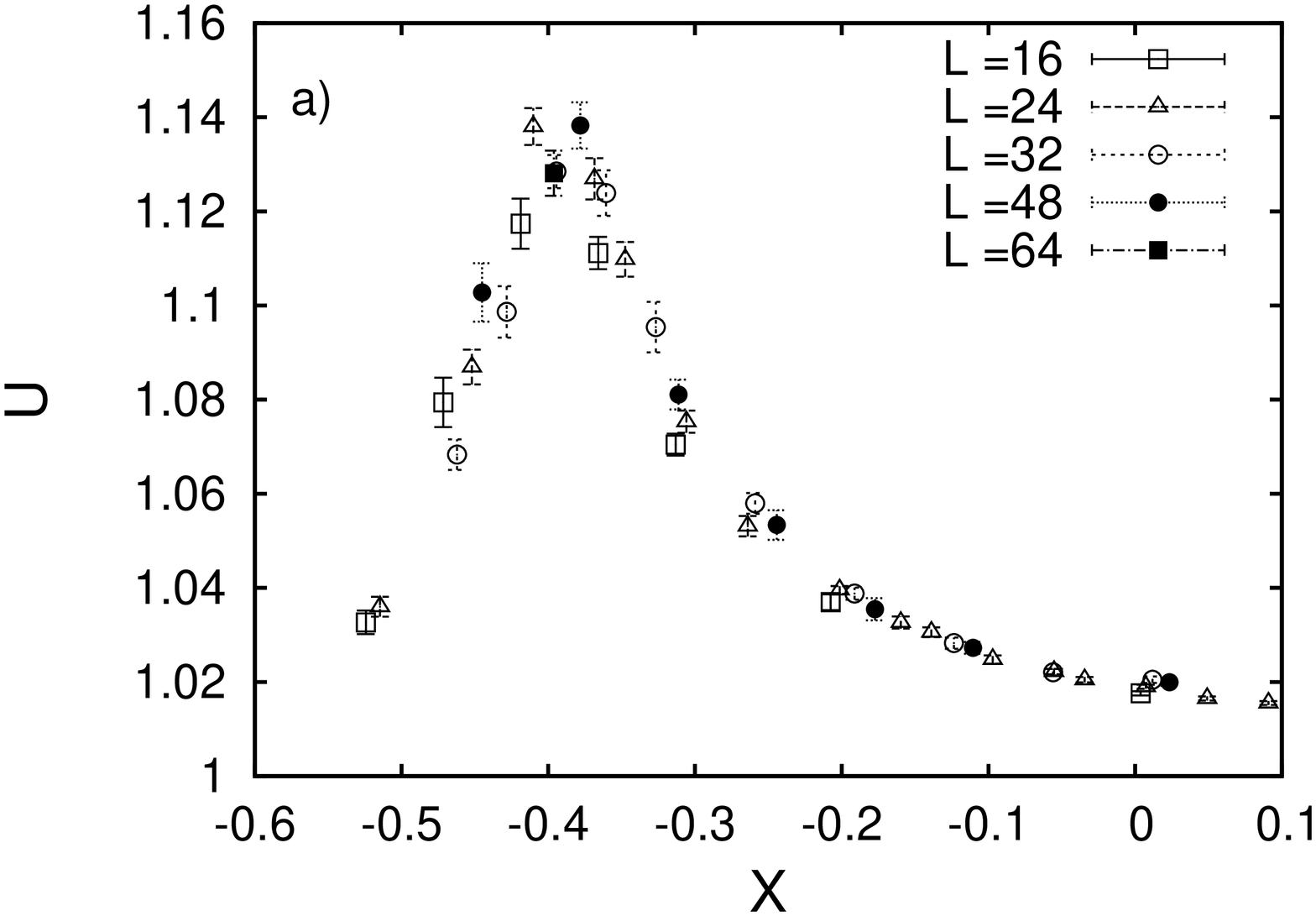}
\includegraphics*[scale=\graphicscale,angle=-90]{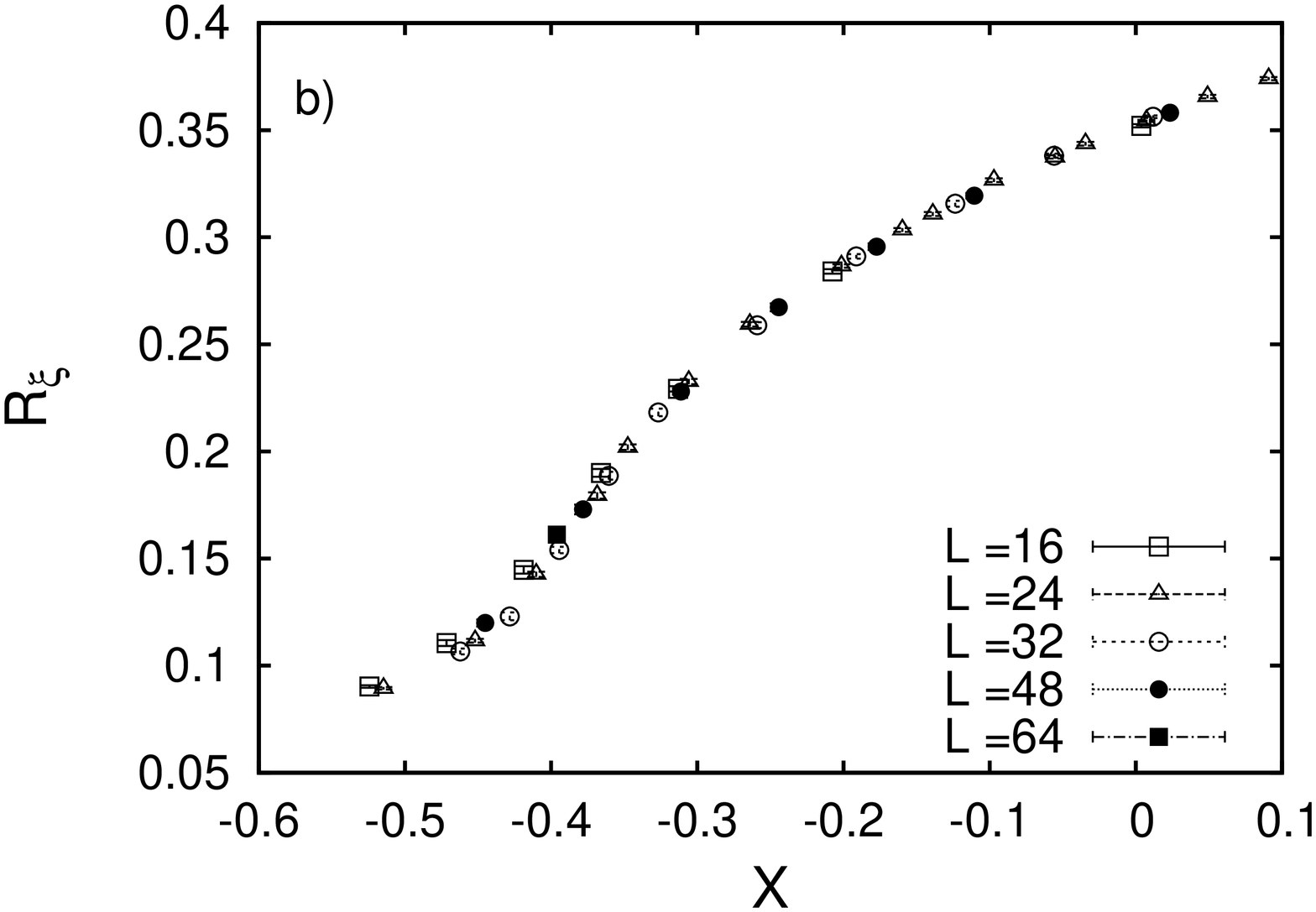}
\caption{Plot of $U$ (top, a) and of $R_{\xi}$ (bottom, b) as a function of 
  $X = (\beta - \beta_c) L^{1/\nu}$, using $\beta_c = 0.319965$ and 
  $\nu = 0.595$. Results for the MFCP$^{24}$ model.
}
\label{scalingplot-N25}
\end{figure}

Next, we determine the critical exponents. We fit the results
for $U$ and $R_\xi$ to Eq.~(\ref{rsca}). We approximate $f_R(x)$ with
a 12th-order polynomial.  The results of the fits are reported in
Table~\ref{fits-nubetac-N25} as a function of $L_{\rm min}$, the
minimum size of the data included in the fit.  There is some
dependence on $L_{\rm min}$, due to scaling corrections.  For $L_{\rm
  min} = 24$, the estimates obtained from $R_\xi$ and $U$ are
consistent, so that we can finally estimate
\begin{equation}
\beta_c = 0.319965(20), \qquad
\nu = 0.595(15).
\label{exponents-N25}
\end{equation}
Errors should be considered as conservative. They are obtained by requiring consistency
between the results obtained for the two values of $L_{\rm min}$. The
data are reported in Fig.~\ref{scalingplot-N25} as a function of $X =
(\beta - \beta_c) L^{1/\nu}$. Scaling is quite good, especially for
the correlation-length ratio. As a consistency check, we have
determined $\nu$ using the specific heat. We find that the the maximum
of the specific heat $C_{\rm max}(L)$ scales as $L^\delta$ with
$\delta = 0.46(15)$. It implies $\nu = 0.58(3)$, which is consistent
with Eq.~(\ref{exponents-N25}).

\begin{figure}[tbp]
\includegraphics*[scale=\graphicscale,angle=-90]{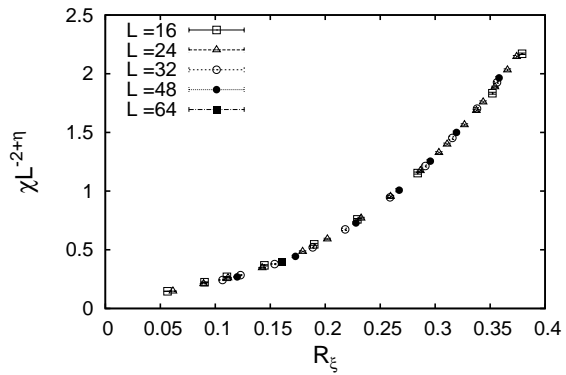}
\caption{Plot of $\chi/L^{2-\eta}$ as a function of $R_\xi$, using
  $\eta = 0.87$. Results for the MFCP$^{24}$ model.  }
\label{scalingplotchi-N25}
\end{figure}

We also study the critical behavior of the susceptibility
$\chi$, performing fits to the ansatz
\begin{equation}
\ln \chi = (2 - \eta) \ln L +
\hat{f}_\chi(R_\xi). 
\end{equation}
 We obtain $\eta = 0.929(3), 0.868(5),
0.871(11)$ for $L_{\rm min} = 16,24,32$.  Note that the results for
the two largest values of $L$ are consistent, allowing us to estimate
\begin{equation}
\eta = 0.87(1).
\end{equation}
In Fig.~\ref{scalingplotchi-N25} we report $\chi/L^{2-\eta} $ versus 
$R_\xi$. The quality of the scaling is excellent.

Finally, we analyze the behavior of the correlation function
$G_V(x)$. We find that $G_V(x)$ behaves as $A \exp(-x/\xi_z)$, where 
the correlation length $\xi_z$ varies between 2.7 and 3.5 in the critical
region for any $L$ in the interval $24\le L \le 64$. Similar results 
are obtained also for $N=4,10,15$. In all cases $\xi_z$ is small: 
we find $\xi_z = 2.1(1), 2.6(1), 2.8(1)$, for  $N=4,10,15$, respectively,
where the error takes into account the $L$ and $\beta$ dependence in the 
transition region. For finite $N$, the correlation $\xi_z$ is expected to 
be finite \cite{DV-80,Aoyama-82,CR-92}. It should, however, diverge in the limit $N\to \infty$, as in
this limit the gauge degrees of freedom are frozen and $\lambda_{{\bm x},\mu}$
can be taken equal to 1. The smallness of $\xi_z$ for $N=25$ indicates that 
we are still quite far from the large-$N$ limit.


\section{Conclusions} \label{sec5}

This paper reports a study of the phase diagram, and the nature of the phase
transitions, of 3D lattice MFCP$^{N-1}$ models characterized by
a global U($N$) symmetry and a U(1) gauge symmetry, and the absence of
monopoles.  We consider the usual lattice nearest-neighbor formulation
of the CP$^{N-1}$ model with an explicit gauge field---the
corresponding Hamiltonian is given in Eq.~(\ref{hcpnla})---restricting
the configuration space to gauge-field configurations in which no
monopoles are present. To define monopoles, we use the definition
proposed by De Grand and Toussaint \cite{DGT-80}.  To determine the 
phase diagram of the 3D MFCP$^{N-1}$ model
we perform Monte Carlo
simulations for $N=2$, 4, 10, 15, and 25.  The analysis of the 
finite-size data allows us to identify a 
finite-temperature transition in all cases, related to the
condensation of a local gauge-invariant bilinear order parameter $Q_{\bm x}$,
cf  Eq.~(\ref{qdef}).

For $N=2$ we considered lattices of size up to $L=80$. In spite of the
relatively large systems considered, we are unable to draw a definite
conclusion on the nature of the transition. We can only safely exclude
that the transition belongs to the O(3) universality class, as it
occurs in the CP$^1$ model in which monopoles are allowed. Some results show
features that are typical of first-order transitions: the results for
the Binder parameter $U$ do not approach a universal curve when
plotted versus $R_\xi = \xi/L$, and the distributions of the order
parameter and of the energy are quite broad, although without the
typical two-peak shape that signals the presence of two coexisting
phases. On the other hand, we do not observe an increase of the
maximum $U_{\rm max}(L)$ of the Binder parameter, which is a signature
of a first-order transition, so that a continuous transition is not
excluded.

If we assume that the MFCP$^1$ has a continuous transition, we
can estimate the critical exponents. For the correlation-length
exponent $\nu$, the quality of the FSS fits is poor. The exponent
significantly decreases as the smallest-volume data are excluded from
the fit, a phenomenon that is often considered as the signature of a
weak first-order transition: in these cases $\nu$ decreases towards
$1/d=1/3$ as larger size data are included.  If we only consider the
largest sizes, we would estimate $\nu = 0.52(2)$, but it is clear from
the quality of the fits that this estimate should be only considered
as an effective estimate in the range of values of $L$ considered.  It
remains an open problem to establish if such a drift stops and the
estimate stabilizes, as appropriate for a continuous transition, or
moves towards the first-order value 1/3.  The exponent $\nu$ can also
be determined from the specific heat, using the hyperscaling relation
$2 - \alpha = 3\nu$.  We obtain $\nu = 0.54(3)$, which is 
consistent with the previous estimate.  We have also analyzed the
behavior of the susceptibility $\chi$ of the order parameter 
$Q_{\bm x}$. In this case, we observe good scaling and little size
dependence of the results. We estimate $\eta = 0.335(10)$. The good
scaling of $\chi$ is presently the only real evidence in favor of a
continuous transition.

It is interesting to compare these results with those obtained in
other models.  An O(3) $\sigma$ model with hedgehog suppression was
considered in Ref.~\cite{MV-04}. The estimates of $\nu$ and $\eta$ are
different from ours, as they obtain $\nu = 1.0(2)$ and $\eta \approx
0.6$. The MFCP$^{1}$ and the model of Ref.~\cite{MV-04} have
the same global symmetry and the same order parameter, but consider
different types of topological defects;  therefore, they may develop a
different behavior.  A loop model, expected~\cite{NCSOS-11,NCSOS-13}
to have the share the same universal large-distance behavior with the MFCP$^1$, was
considered in Ref.~\cite{NCSOS-15}. The numerical results obtained on very large
systems (up to $L=512$) show some similarities, but also some notable
differences, with ours.  For instance, they also find significant
violations of FSS and a significant dependence of the estimates of
$\nu$ from the system sizes considered. The loop-model estimates of
$\nu$ vary from 0.6 at small sizes to $\nu\approx 0.46$ for the
largest ones. We can also compare our estimate of $\eta$ with that
obtained in Ref.~\cite{NCSOS-15} for the N\'eel order parameter, which
corresponds to our operator $Q_{\bm x}$. The FSS analysis of the order
parameter or of the corresponding susceptibility shows significant FSS
violations. On the other hand, the analysis
of the short-distance behavior of the two-point correlation 
function gives a quite clear power-law behavior up to distances $r\sim 100$
with quite good scaling collapse. This allows Ref.~\cite{NCSOS-15} to 
estimate $\eta = 0.259(6)$,
which, however, significantly differs from our estimate $\eta = 0.335(10)$. 
This difference may be explained either by a different universality class
or by the fact that at least one of
the two models does not undergo a continuous transition. Of course, it
is also possible, as suggested in the literature on quantum
antiferromagnets, that the transition is continuous with anomalously
large and slowly decaying---even logarithmic \cite{Sandvik-10}, 
associated with a
dangerously irrelevant variable---scaling corrections. 

The results for $N=4,10$, and 15 are instead quite conclusive on the
order of the transition. In all cases, we have clear evidence that the
transition is of first order. The maximum $U_{\rm max}(L)$ of the
Binder parameter increases with $L$ and, for sufficiently large $L$,
we observe two maxima in the distributions of the order parameter and
of the energy (for the energy only for $N=10,15$). The transition is
significantly weaker than in the usual CP$^{N-1}$ model in which
monopoles are allowed. In particular, while in the latter model the
transition becomes stronger as $N$ increases, for the MFCP$^{N-1}$ the
opposite occurs for $N\ge 10$: the transition for $N=10$ is stronger
than for $N=15$. As expected for first-order transitions, the maximum
$C_{\rm max}(L)$ of the specific heat diverges. In our range of values
of $L$, however, the increase is slower than the expected one, $C_{\rm
  max}(L)\sim L^d = L^3$. Apparently, it increases as $L^\delta$, with
$\delta \sim 1$, which would imply, using the usual relations valid
for continuous transitions, an effective exponent $\nu$ of the order
of 0.5. This shows that effective estimates of $\nu$ around 1/2 are not uncommon
in the presence of weak first-order transitions, casting additional
doubts on the interpretation of the results for $N=2$ as a continuous
transition.

Our conclusions for the nature of the transition for $4\le N \le 15$
differ from those of Refs.~\cite{KS-12,BMK-13}, that observed instead
continuous transitions in the same range of values of $N$. Note,
however, that this is not necessarily an inconsistency. A priori, it
is always possible that a MFCP$^{N-1}$ fixed point exists for these values 
of $N$, but that our model is outside its attraction domain.
We mention that the existence of a range of value of $N$, where the model
undergoes a weakly first-order transition is consistent with 
the renormalization-group analysis of Ref.~\cite{IZMHS-19}.

Finally, we have studied the MFCP$^{N-1}$ model with $N=25$. In this case,
data are consistent with a conventional continuous transition. 
Data (with $L$ up to 64) show a good FSS, with exponents 
\begin{equation}
   \nu = 0.595(15) ,\qquad \eta = 0.87(1).
\end{equation}

\begin{table}
\caption{Estimates of the transition point for the CP$^{N-1}$ model with 
monopoles ($\beta_c^{std}$), taken from Refs.~\cite{PV-19-CP,PV-20-largeN},
and for the MFCP$^{N-1}$ model without monopoles ($\beta_c$). }
\label{betac-table}
\begin{tabular}{ccc}
\hline\hline
$N$ & $\beta_c$ & $\beta_c^{\rm std}$ \\
\hline
2   & 0.4605(3)   & 0.7102(1) \\
4   & 0.4285(5)   & 0.5636(1) \\
10  & 0.3712(3)   & 0.4253(5) \\ 
15  & 0.3472(3)   & 0.381(1)  \\
20  &             & 0.353(2)  \\
25  & 0.319965(20)&   \\
\hline\hline
\end{tabular}
\end{table}

It is interesting to compare our results with the predictions of the
field-theory approaches that are used to describe the large-distance
behavior of the model: the gauge-invariant Landau-Ginzburg-Wilson
(LGW) approach, see Ref.~\cite{PV-19-CP,PV-19-AH3d}, which has also
been successfully applied to systems with nonabelian gauge symmetries
\cite{PW-84,BPV-nonabelian}, and the Abelian-Higgs field theory
\cite{HLM-74,FH-96,MZ-03,IZMHS-19}.
The first approach predicts a first-order transition for $N\ge 3$.
For $N=2$, continuous transitions necessarily belong to the O(3)
universality class. Our results are clearly not consistent with the
LGW predictions, as we find a continuous transition for $N=25$ (the
results for $N=2$ might still be consistent with the LGW approach if
the phase transition of the MFCP$^{1}$ model is of first order).
This shows that the LGW approach is not appropriate to describe the
monopole-free model. However, this is not surprising. If monopoles are
relevant in defining the long-distance behavior of the model, the
effective theory should include somehow the information on the
topology of the gauge fields. This is clearly not possible in the LGW
approach, as the gauge degrees of freedom are integrated out.

The Abelian-Higgs field theory \cite{HLM-74} predicts continuous
transitions for $N> N_{c,FT}$ and first-order ones for $N< N_{c,FT}$.
Close to 4 dimensions, we have \cite{HLM-74} $N_{c,FT}\approx 183$.  A
three-dimensional estimate is quite problematic to obtain, because of
the non-Borel summability of the perturbative series in powers of
$\epsilon = 4 - d$. Ref.~\cite{IZMHS-19} quotes $N_{c,FT} = 12.2
(3.9)$.  It is tempting to conjecture that the continuous transition
we have observed for $N=25$ is associated with the stable large-$N$
fixed point occurring in the Abelian-Higgs field theory. This would
also be supported by the fact that our estimate of the value $N_c$
separating first-order from continuous transitions, which should
belong to the interval $15 < N_c < 25$, is essentially consistent with
the field-theory estimate of Ref.~\cite{IZMHS-19}. 

Finally, let us consider the behavior for $N\to\infty$. In
Ref.~\cite{PV-20-largeN}, we showed that the model with Hamiltonian
(\ref{hcpnla}) has a first-order transition for any $N\ge 3$,
including $N=\infty$, contradicting the analytic computations of
Ref.~\cite{DHMNP-81}.  It was conjectured that the failure is due to
the presence of monopoles in the disordered phase that do not allow
the ordering of the gauge fields \cite{MS-90}, even for $N=\infty$.
If this interpretation is correct, the MFCP$^{N-1}$ model should
instead give results consistent with the analytic computations of
Ref.~\cite{DHMNP-81} in the large-$N$ limit. The fact that the
transition becomes continuous as $N$ increases supports this
conjecture. A more quantitative check can be performed using the
large-$N$ estimates~\cite{HLM-74,YKK-96,KS-08}
\begin{equation}
\eta = 1 - {32\over \pi^2 N} ,\qquad
\nu = 1 - {48\over \pi^2 N} .
\end{equation}
For $N=25$ they give $\eta = 0.87$ and $\nu = 0.81$. The estimate of
$\eta$ is in perfect agreement with our result, while the estimate of
$\nu$ differs considerably. This is, however, not totally surprising,
since the critical value $N_c$ where the order of the transition
changes (consequently $1/N_c$ is expected to be the radius of the
region in which the large-$N$ expansion is predictive) may be close to
25. If this occurs, it is clear that a quantitative agreement requires
considering several terms of the expansion. As a final remark, we note
that the difference $\beta_c(N)^{\rm std} - \beta_c(N)$
($\beta_c(N)^{\rm std}$ and $\beta_c(N)$ are the transition points for
the model with and without monopoles, respectively, reported in
Table~\ref{betac-table}) scales quite precisely as $1/N$. This leads
us to conjecture that, for $N=\infty$, monopoles do not change the
transition temperature, but only the nature of the disordered
high-temperature phase.


\begin{thebibliography}{99}
\bibitem{footnoteMC} To update each lattice variable, we randomly
  choose either a standard Metropolis update, which ensures
  ergodicity, or a microcanonical move, which is more efficient than
  the Metropolis one but does not change the energy. Typically, on
  average we perform four/three microcanonical updates for every
  Metropolis proposal.  In the Metropolis update, changes are tuned so
  that the acceptance is approximately 1/3.
\bibitem{footnoteO3} We have performed simulations in the usual 
O(3) vector model with periodic boundary conditions to determine 
the FSS curve of the Binder parameter $U = F(R_\xi)$. An interpolation is 
$F(x) = 1.666666 + x (3.0263535 + 23.139470 x)(1-e^{-15x}) 
- 47.838890 x^2  + 58.489668 x^3  - 67.020681 x^4  + 38.408855 x^5
- 8.8557348 x^6$. The error is smaller than 0.5\%.
\end{thebibliography}
\end{document}